\documentclass[useamsfonts]{pasj01}
\usepackage{graphicx}

\begin{document}

\title{Search for gas bulk motions in eight nearby clusters of galaxies with Suzaku}

\author{Naomi \textsc{Ota}\altaffilmark{1}}
\email{naomi@cc.nara-wu.ac.jp}

\author{Hiroko \textsc{Yoshida}\altaffilmark{1}}

\altaffiltext{1}{Department of Physics, Nara Women's University, Kitauoyanishi-machi, Nara, Nara 630-8506, Japan}

\maketitle

\begin{abstract}
  To search for bulk motions of the intracluster medium, we analyzed
  the X-ray spectra taken with the {\it Suzaku} satellite and measured
  the Doppler shift of Fe-K line emission from eight nearby clusters
  of galaxies with various X-ray morphologies. In the cores of the
  Centaurus and Perseus clusters, the gas bulk velocity does not
  exceed the sound velocity, which confirms the results of previous
  research. For the Cen45 subcluster, we found that the radial
  velocity relative to the Centaurus core, $<780~{\rm km\,s^{-1}}$, is
  significantly smaller than that reported in the optical band at
    the $3.9\sigma$ level, which suggests an offset between the gas
  and galaxy distributions along the line of sight due to the
    subcluster merger. In A2199, A2142, A3667, and A133, no
  significant bulk motion was detected, indicating an upper limit on
  the radial velocity of $3000-4000~{\rm km\,s^{-1}}$. A sign of
    large bulk velocity in excess of the instrumental calibration
    uncertainty was found near the center of cool-core cluster A2029
    and in the subcluster of the merging cluster A2255, suggesting
    that the nonthermal pressure support is not negligible in
    estimating the total gravitational mass of not only merging
    clusters but also relaxed clusters as predicted by numerical
    simulations. To improve the significance of the detection,
    however, a further examination by follow-up observations is
    required. The present study provides a pilot survey prior to the
  future high-resolution spectroscopy with {\it ASTRO-H}, which is
  expected to play a critical role in revealing the dynamical
  evolutions of clusters.
  \end{abstract}

\section{Introduction}
Galaxy clusters comprise the largest structures in the universe and
the precise knowledge of their mass structures is important for
understanding the history of structure formation and to constrain the
still unknown types of dark content of the universe. The clusters are
believed to have grown into their present shapes via collisions and
mergers of smaller groups and clusters. If such objects collide under
gravitational attraction, a huge amount of energy can possibly be
released. A certain fraction of this energy is expected to heat the
intracluster medium (ICM), generate nonthermal particles through shock
waves, and induce bulk and turbulent gas motions. Earlier simulations
have predicted that the gas bulk motion with the speed of a
substantial fraction of virial velocity ($>1000~{\rm km\,s^{-1}}$)
will persist for several giga years (e.g.,
\cite{1997ApJS..109..307R,1999LNP...530..106N,2001ApJ...561..621R,2005MNRAS.364..753D}).

If the kinetic velocity of the gas is larger than its sound velocity,
nonthermal pressure cannot be neglected in estimating the total
gravitational mass inside the cluster.  \citet{2007ApJ...655...98N}
used numerical simulations to show that the hydrostatic mass estimate
is biased low by approximately 5--20\% even in relaxed clusters. This
result can be primarily attributed to an additional pressure support
provided by the bulk motions in the ICM. The departure from the
hydrostatic assumption is one of the largest sources of systematic
uncertainties in cluster cosmology (e.g.,
\cite{2009ApJ...705.1129L,2012NJPh...14e5018R,2013SSRv..177..119E}).

The bulk and turbulence gas motions produce Doppler shifts in emission
lines from heavy irons in the ICM and modify the line profiles
\citep{2003AstL...29..791I}. A line-of-sight velocity of $1000~{\rm
  km\,s^{-1}}$ corresponds to a 22~eV shift in the 6.7 keV Fe-K line
energy. Therefore the Doppler-shift measurement requires not only a
high energy resolution and a good sensitivity but also a precise
instrumental energy-scale calibration.

Observational efforts that used X-ray charge-coupled device (CCD)
detectors with a typical energy resolution of 150~eV have constrained
gas motion (for review, \cite{2012RAA....12..973O}). On the basis of
careful assessment of the energy scale of the X-ray Imaging
Spectrometer (XIS; \cite{2007PASJ...59S..23K}) onboard the {\it
  Suzaku} satellite \citep{2007PASJ...59S...1M},
\citet{2007PASJ...59S.351O} placed an upper limit of $1400~{\rm
  km\,s^{-1}}$ on bulk motions in the Centaurus cluster and suggested
that the cluster mass estimated under the hydrostatic assumption is
justified within a factor of 2--3. Similar measurements that used XIS
data have been conducted in several nearby clusters: the upper limit
on the bulk motion was derived for the Ophiuchus cluster, AWM7, A2319,
the Coma cluster, and A3627 \citep{2008PASJ...60.1133F,
  2008PASJ...60S.333S,2009PASJ...61.1293S, 2011PASJ...63S.991S,
  2012PASJ...64...16N}, and possible detection of gas bulk flow in
Centaurus and A576 was reported by {\it Chandra}
\citep{2007ApJ...668..781D,2006ApJ...639..781D}. Thus far, bulk
motions has been properly detected in only one cluster: a large radial
velocity of a subcluster region relative to the main cluster,
$1500~{\rm km\,s^{-1}}$, was found in A2256 by {\it Suzaku}
\citep{2011PASJ...63S1009T}. They suggested that the gas and galaxies
are moving together in the cluster potential as a single
subcluster. Doppler line broadening has been measured by the
Reflection Grating Spectrometer on {\it XMM-Newton}, and
\citet{2010MNRAS.402L..11S} obtained a strong upper limit on the
turbulent motion ($<200~{\rm km\,s^{-1}}$) in several objects.
\citet{2015A&A...575A..38P} measured the turbulent line broadening for the RGS cluster
sample and reported that upper limits generally range between
$200-600~{\rm km\,s^{-1}}$ in the 0'.8 and 3'.4 core regions.

These previous studies demonstrated that the {\it Suzaku}/XIS
spectroscopy enables meaningful constraints to be placed on bulk
velocity distribution and is complementary to the turbulence
measurements by the gratings, which can be performed only at the
innermost region.  The number of sample, however, is still
insufficient for clarifying the detailed velocity structure of the ICM
and for understanding the role of kinetic gas motion in the precise
cluster mass estimation. In this study, we perform a systematic search
for gas bulk motion in nearby clusters by utilizing the {\it Suzaku}
archive data. The sample includes eight clusters in the nearby
universe that have a wide range of X-ray morphology. We aim to
constrain the presence of gas bulk motions and to determine the
dynamical state of clusters by comparing the velocity structure of the
ICM with that of member galaxies in the optical band. This study
provides an important pilot survey prior to the {\it ASTRO-H} mission
\citep{2014SPIE.9144E..25T} because it ranks the investigation of
turbulent and macroscopic gas motions as main objectives of
high-resolution spectroscopy \citep{2014SPIE.9144E..2AM}.

Throughout this study, we adopt $\Omega_M=0.27$,
$\Omega_{\Lambda}=0.73$, and $H_0 = 70~{\rm km\,s^{-1}Mpc^{-1}}$. The
quoted statistical errors refer to the 68\% confidence ranges, unless
stated otherwise.

\section{Observation and data reduction}
To investigate the ICM velocity structure and its possible
relationship with the X-ray morphology, the sample was selected to
include clusters with cool cores, cold fronts, and subcluster
mergers. The observational features of individual clusters will be
mentioned in detail in \S\ref{subsec:velocity_structure}.  Because a
measurement of the Doppler shift of the Fe-K emission line requires a
good photon statistics, the sample was limited to nearby objects
  with the 2--10~keV X-ray flux $\gtrsim 10^{-11}~{\rm
    erg\,s^{-1}\,cm^{-2}}$ within the {\it Suzaku} data archive.  As
shown in Table~\ref{tab1}, the sample consists of eight nearby
clusters including the Centaurus and Perseus clusters, A2199, A3667,
A133, A2029, A2255, and A2142. The redshift values of the target
clusters, $z_{\rm cl}$, were cited from the NASA/IPAC Extragalactic
Database, except for the Centaurus cluster ($z=0.0104$;
\cite{1986MNRAS.220..679D}) and A2142 ($z=0.0909$;
\cite{2011ApJ...741..122O}).

\begin{table*}
\tbl{Log of Suzaku observations of eight clusters}{ 
\begin{tabular}{lllllllll} \hline 
Pointing & Obsid& Date & $z_{\rm cl}$ & RA & Dec &Exp$^{\mathrm{a}}$  & BM$^{\mathrm{b}}$ & kpc/1\arcmin \\
 & &  &  & (deg) & (deg) & (ks)  & &     \\ \hline
Centaurus  Center& 800014010 & 2005 Dec 27 &  0.0104 & 192.2012 & $-41.3132$  & 35.8 & I--I\hspace{-.1em}I & 13\\
Centaurus S & 800015010 & 2005 Dec 28 & --  & 192.2015 & $-41.4461$ & 42.6   & -- & -- \\ 
Centaurus N & 800016010 & 2005 Dec 29 & -- & 192.2009 & $-41.1804$ & 42.0   &-- & -- \\ 
Cen45 & 802008010 & 2007 Dec 24 & -- & 192.5119 & $-41.3865$ & 57.5  & -- & -- \\ \hline
Perseus &101012010 & 2006 Aug 29 & 0.0183 & 49.9436 & 41.5175 & 48.8 & I\hspace{-.1em}I--I\hspace{-.1em}I\hspace{-.1em}I & 22 \\ \hline
A2199 Center  & 801056010 & 2006 Oct 01  & 0.0302 & 247.1922 & 39.4840 & 25.1  & I & 36 \\ 
A2199 SE  & 801057010 & 2006 Oct 01 & -- & 247.4695 & 38.9998 & 23.7 &  -- & --  \\ 
A2199 SW   & 801058010 & 2006 Oct 03 & -- & 247.0222 & 38.9947 & 22.4 & -- & --  \\ 
A2199 NW  & 801059010 & 2006 Oct 04  &  -- & 246.6242 & 39.5648 & 24.7 & -- & -- \\ 
A2199 NE   & 801060010 & 2006 Oct 04  &  -- & 247.7662 & 39.3698 & 24.3 & -- & -- \\ \hline 
A3667 Center & 801096010& 2006 May 06 & 0.0556 & 303.1410 & $-56.7974$ & 20.6   & I--I\hspace{-.1em}I & 65 \\
A3667 NW 1 & 801095010 & 2006 May 06 & -- & 302.8177 & $-56.5600$ & 17.1  & -- & -- \\
A3667 NW 2 & 801095020 & 2006 Oct 30 & -- & 302.8690 & $-56.6666$ &11.5 & -- &-- \\ 
A3667 SE &805036010 & 2010 Apr 12 & -- & 303.4605 & $-57.0338$ & 64.2  & -- & -- \\ \hline
A133 W & 805019010& 2010 Jun 07 & 0.0566 & 15.4513 & $-21.8877$ & 48.5   & -- & 66 \\ 
A133 N & 805020010& 2010 Jun 05 & -- & 15.6732 & $-21.6726$ &49.0  & -- & -- \\
A133 E & 805021010& 2010 Jun 09 & -- & 15.9075 & $-21.8935$ &50.8  & -- & -- \\
A133 S & 805022010 & 2010 Jun 08 & -- & 15.6819 & $-22.1029$ &50.2  & -- & -- \\ \hline
A2029 Center & 804024010 & 2010 Jan 28 & 0.0773 & 227.7409 & 5.7498 & 7.6 & I  & 88 \\ 
A2029 N &804024020  & 2010 Jan 28 & -- & 227.8527 & 6.0108 & 25.1 & -- & -- \\ 
A2029 S & 804024030 &  2010 Jan 28 & -- & 227.6311 & 5.4874 & 17.3 & -- & -- \\ 
A2029 E & 804024040 &  2010 Jan 29 &  -- & 228.0053 & 5.6496 & 24.9  & -- & -- \\ 
A2029 W & 804024050 &  2010 Jan 30 & -- & 227.4808 & 5.8608 & 21.4   & -- & -- \\ \hline 
A2255 Center & 804041010& 2010 Feb 07 & 0.0806 & 258.2484 & 64.1467 & 44.0  & I\hspace{-.1em}I--I\hspace{-.1em}I\hspace{-.1em}I & 91 \\ 
A2255 NW & 808039010 & 2013 Nov 17 & -- & 258.0206  & 64.3236 & 42.3  & --  & -- \\ \hline
A2142 & 801055010& 2007 Jan 04 & 0.0909 & 239.5312 & 27.2879 & 50.8  & I\hspace{-.1em}I &102 \\ \hline
\end{tabular}}\label{tab1}
\begin{tabnote}
$^{\mathrm{a}}$ Net exposure time of the XIS sensors after data filtering. $^{\mathrm{b}}$ Bautz-Morgan type \citep{1970ApJ...162L.149B}.
\end{tabnote}
\end{table*}

We utilised X-ray spectral data obtained with XIS onboard the {\it
  Suzaku} satellite because it has an excellent sensitivity at the
Fe-K line energies and the lowest background. The XIS consists of four
X-ray CCD cameras including three front-illuminated (FI) CCDs (XIS-0,
XIS-2\footnote{XIS-2 is not operational as of November 2006.} , XIS-3)
and one back-illuminated (BI) CCD (XIS-1). The XIS was operated in
normal modes during the observations.  The details of the observations
are given in Table \ref{tab1}, and the XIS images are shown in
Fig.~\ref{fig1}.

Data reduction was performed by using HEASOFT version 6.16, together
with the calibration database released on October 1, 2014. We
extracted light curves to confirm that the count rates are constant
within the photon statistics. As we will detail in
\S\ref{sec:analysis}, spectral analysis was performed in two types of
energy ranges including 0.7--10~keV and 5--10~keV. For the 0.7--10~keV band, the background was estimated from the blank-sky
data obtained during the Lockman Hole observations, whereas the
background for the 5--10~keV band is dominated by an instrumental
non-X-ray background and was thus calculated using the {\tt xisnxbgen}
tool in the HEASOFT package. The energy response files were generated
by using {\tt xisrmfgen}, and the auxiliary response files were
calculated using {\tt xissimarfgen} \citep{2007PASJ...59S.113I}, in
which the X-ray surface brightness of each cluster was modeled by the
$\beta$-model \citep{1999ApJ...517..627M, 2006ApJ...648..176L}.

The corners of the XIS CCD cameras illuminated by $^{55}$Fe
calibration sources, as well as circular regions with radius
$1\arcmin$ centered on point sources detected in the {\it XMM-Newton}
images were excluded from spectral integration regions. Here the {\it
  XMM-Newton} MOS data were reduced through the standard manner by
using Science Analysis System (SAS) version 13.0.0. The event files
were created by using the {\tt emchain} tool in the SAS package.  We
confirmed that the contribution from the point sources was negligible
in our Fe-line analysis by changing the extraction radii between
0'--1'.5.

\section{Spectral analysis and results}\label{sec:analysis}
To study the dynamical state of ICM, we measured the redshift of Fe-K
emission lines for eight clusters of galaxies. The definition of
spectral regions is described in \S\ref{subsec:region}. The method and
results of spectral fitting are shown in
\S\ref{subsec:fitting}. Because the accurate calibration of the energy
scale is critical for the present purpose, we estimated the systematic
error associated with the spectral analysis in \S\ref{subsec:accuracy}
and constrained the velocity of ICM bulk motions in
\S\ref{subsec:bulk_constraint}.

\subsection{Definition of spectral regions}\label{subsec:region}
The analysis consisted of two steps: (i) We first measured the
redshift of the Fe-K lines, $z$, by using X-ray spectra accumulated
from the entire $18\arcmin\times18\arcmin$ XIS field of view (FOV) to
infer the mean radial velocity of the ICM. (ii) We further divided the
FOV into smaller cells to constrain the spatial velocity distribution
if the cluster emission was bright enough to obtain sufficient photon
statistics in each cell. Although the actual physical scales of the
cells may differ depending on the distance to the object, we hereafter
refer to the above two cases simply as (i) the large-scale
($18\arcmin$) and (ii) the small-scale ($4\arcmin.5-9\arcmin.0$)
velocity measurements.

\begin{figure*}
\begin{center}
\rotatebox{0}{\scalebox{0.24}{\includegraphics{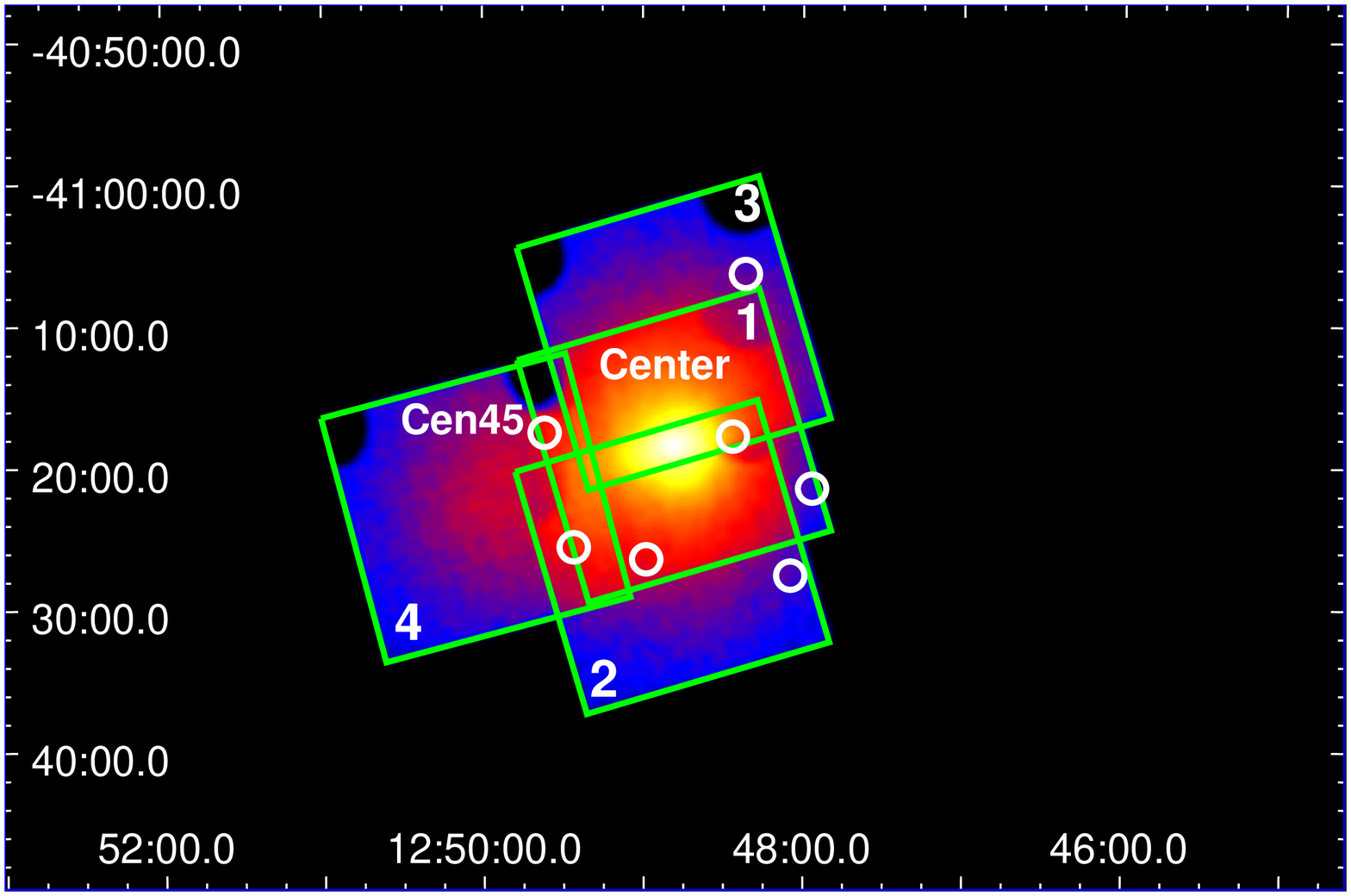}}}
\rotatebox{0}{\scalebox{0.25}{\includegraphics{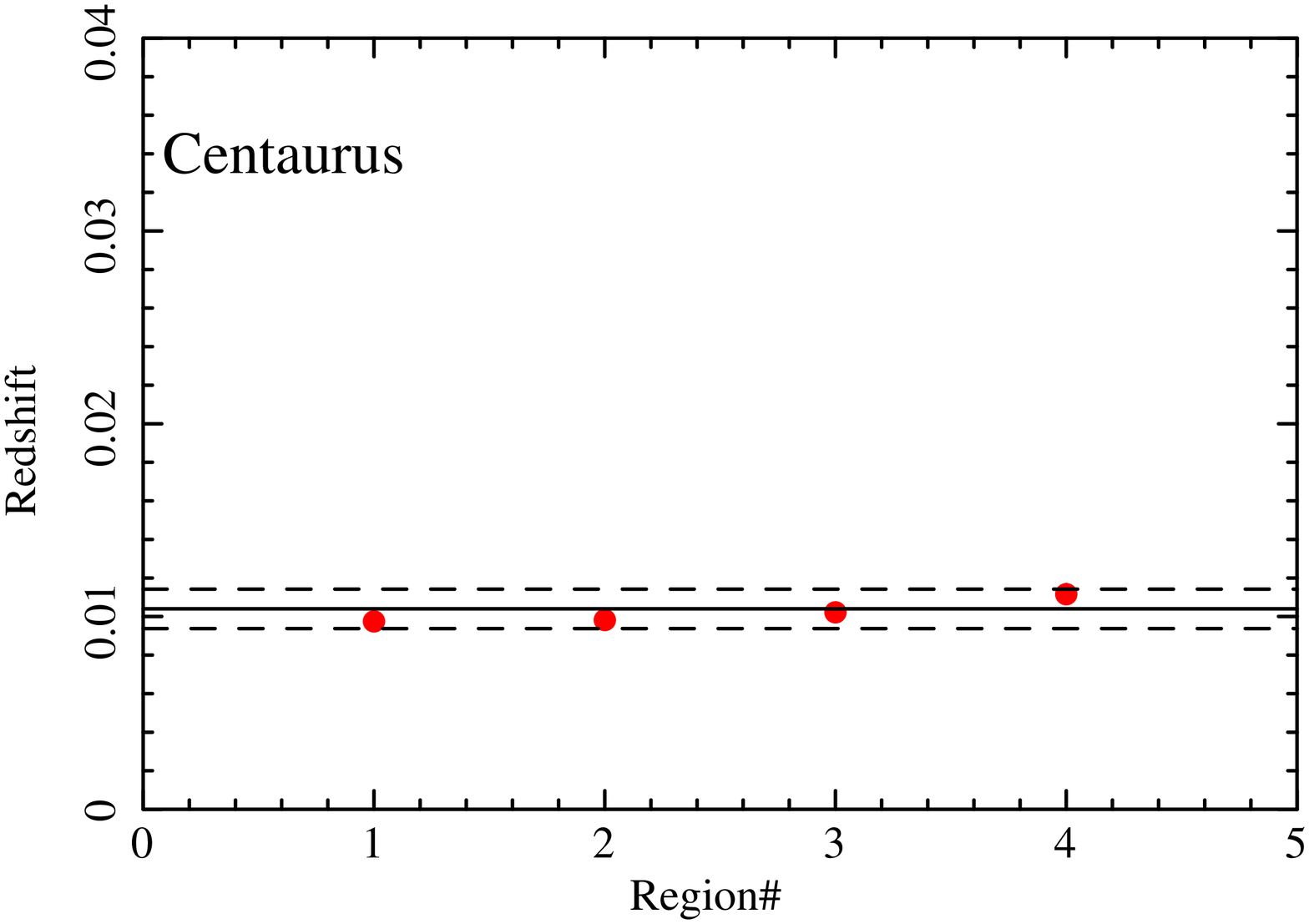}}}

\rotatebox{0}{\scalebox{0.24}{\includegraphics{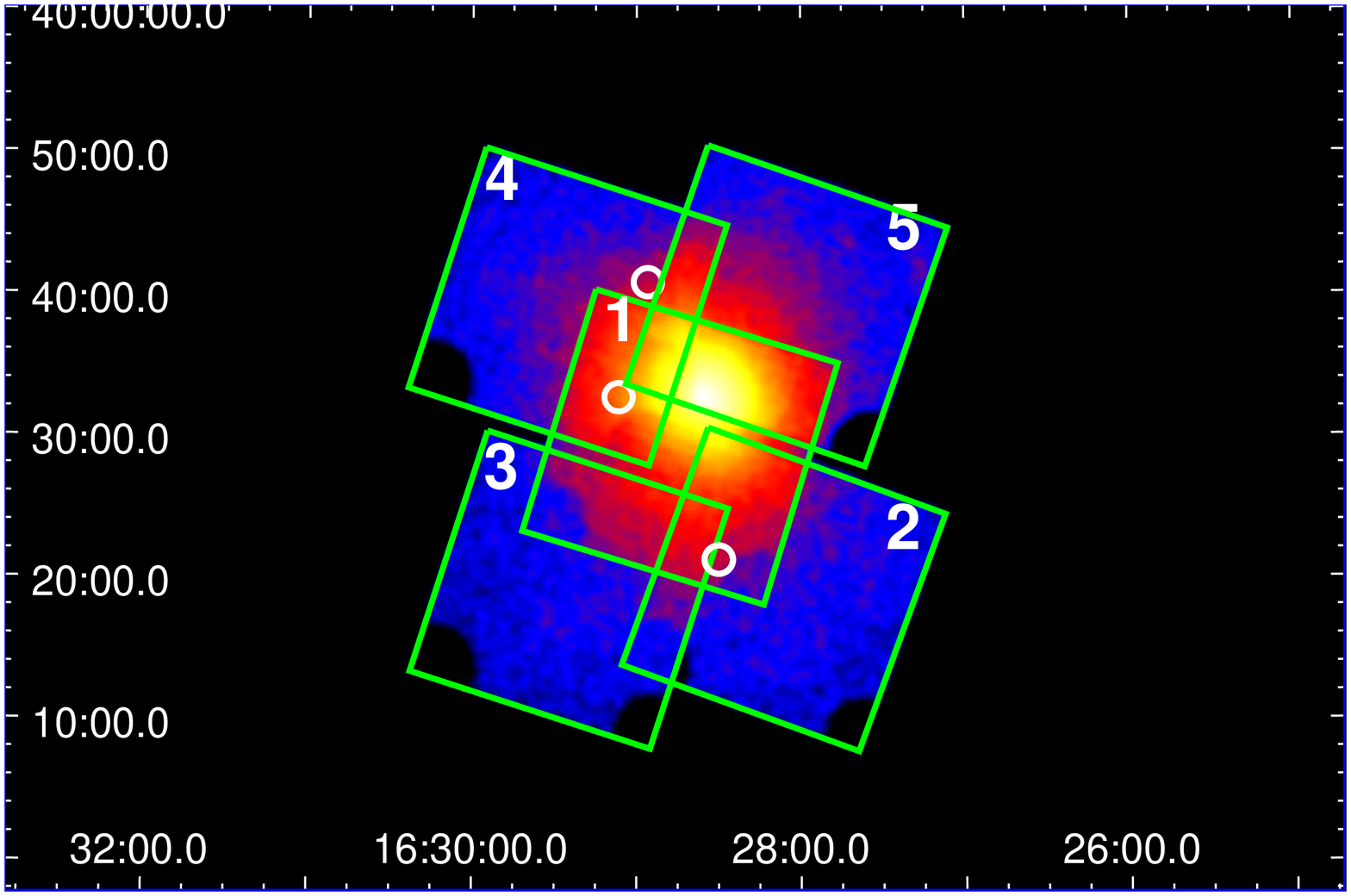}}}
\rotatebox{0}{\scalebox{0.25}{\includegraphics{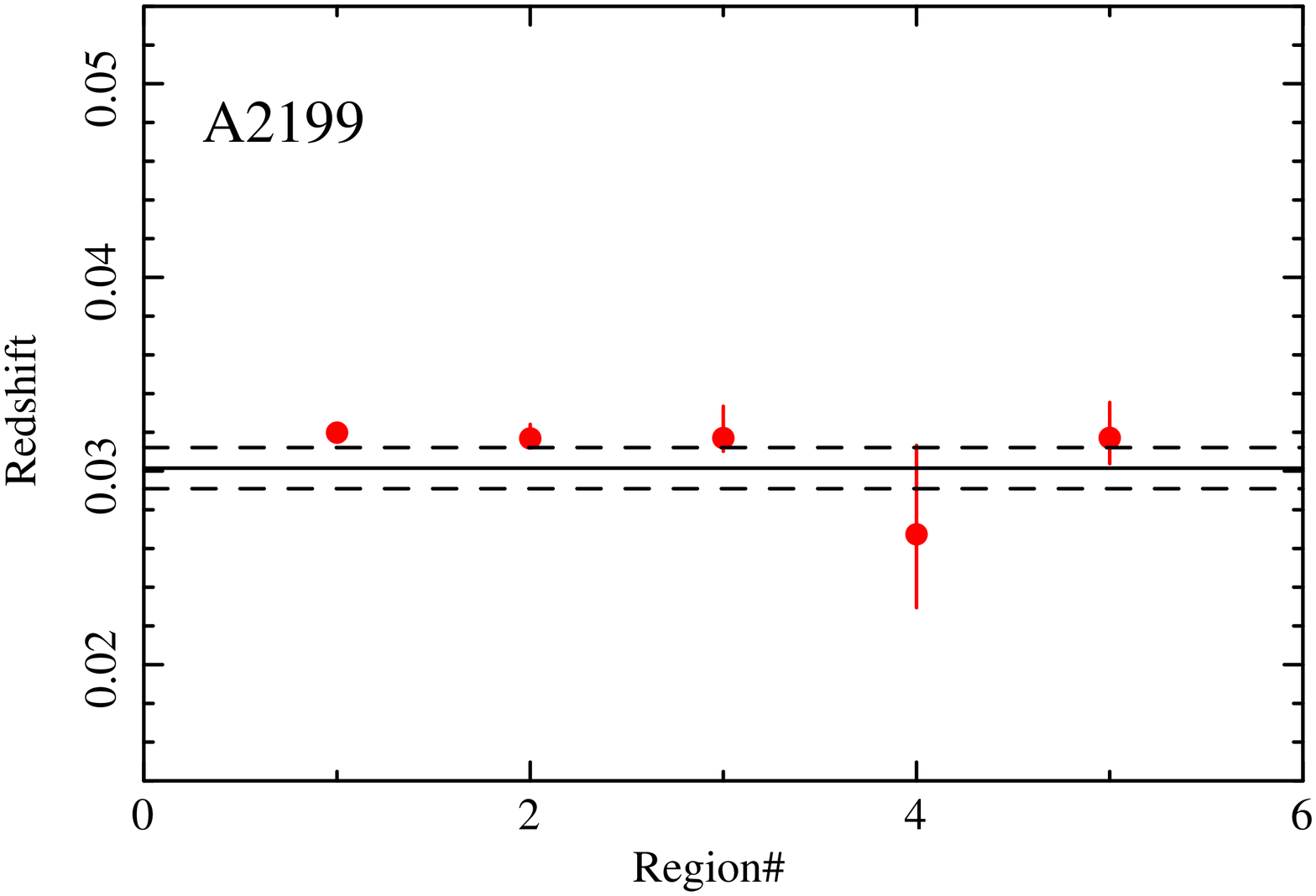}}}

\rotatebox{0}{\scalebox{0.24}{\includegraphics{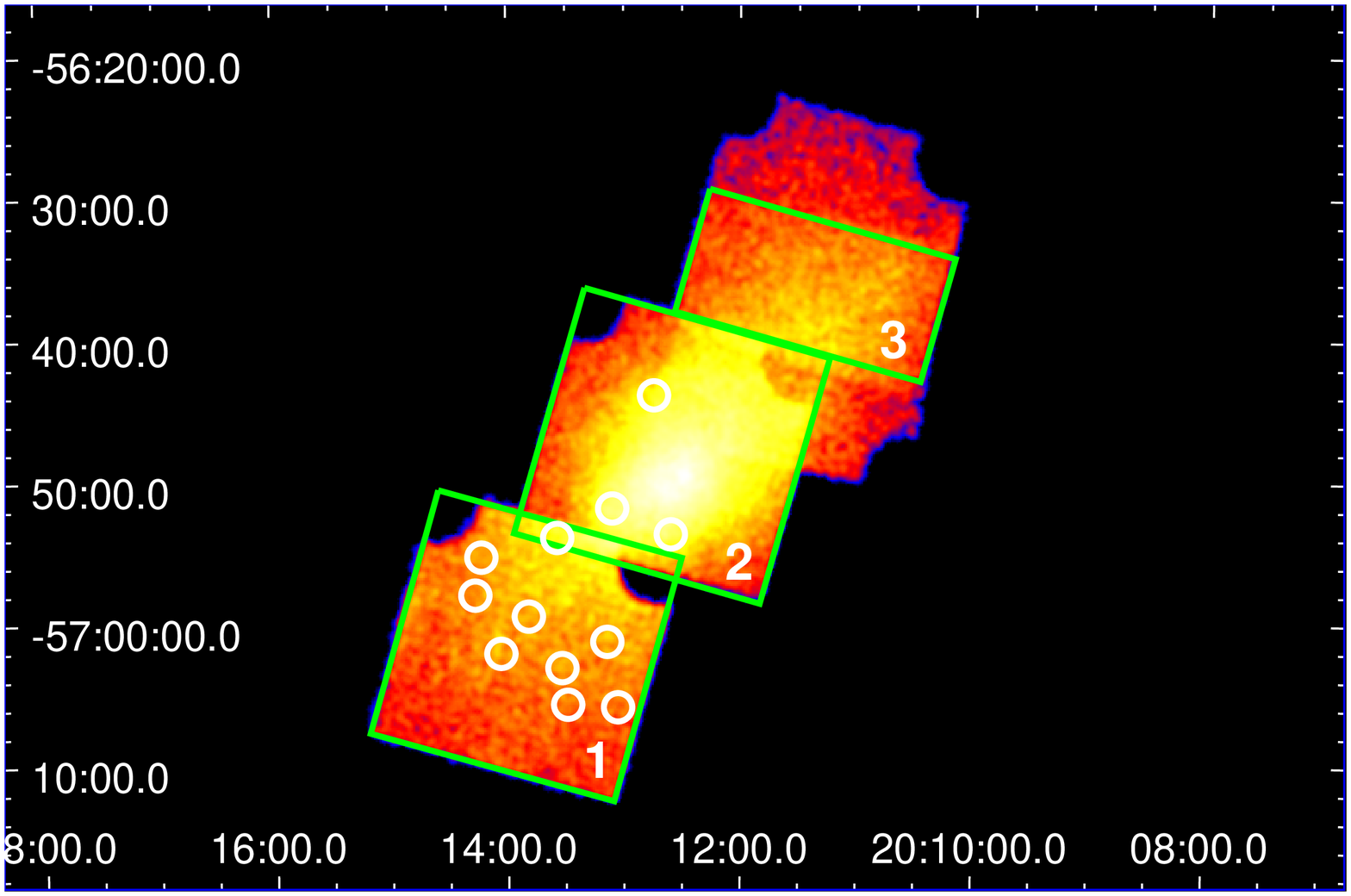}}}
\rotatebox{0}{\scalebox{0.25}{\includegraphics{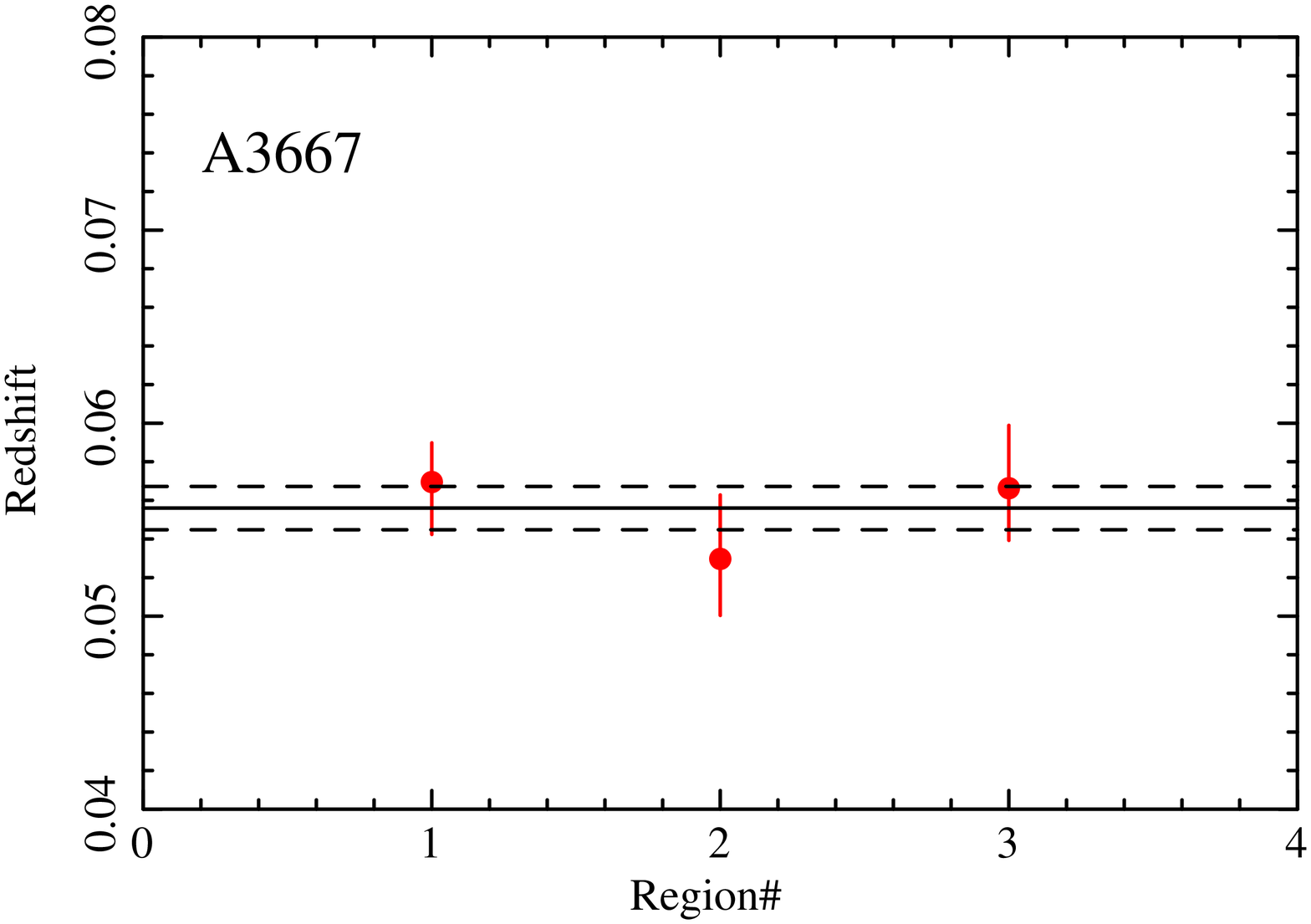}}}

\rotatebox{0}{\scalebox{0.24}{\includegraphics{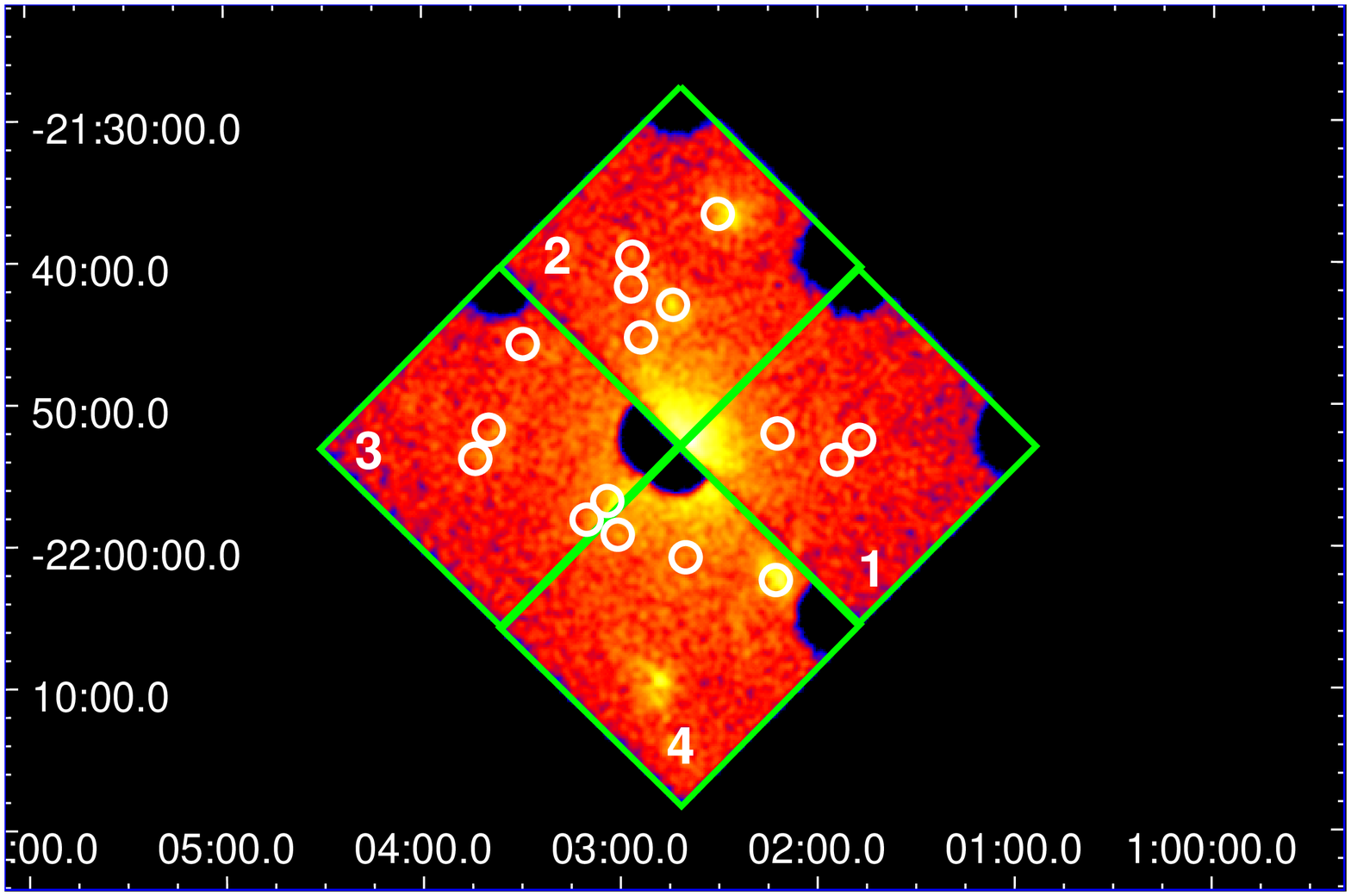}}}
\rotatebox{0}{\scalebox{0.25}{\includegraphics{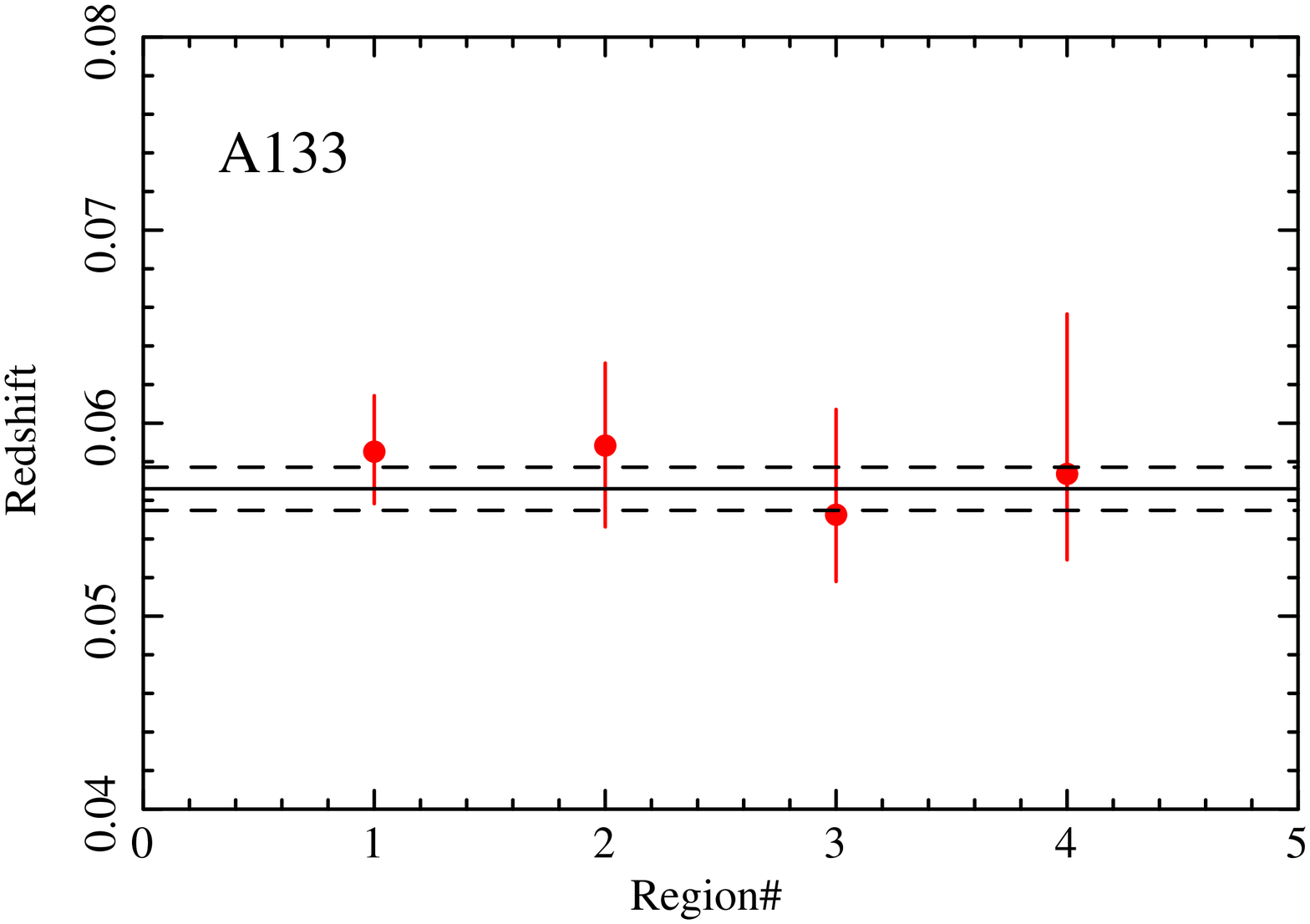}}}

\rotatebox{0}{\scalebox{0.24}{\includegraphics{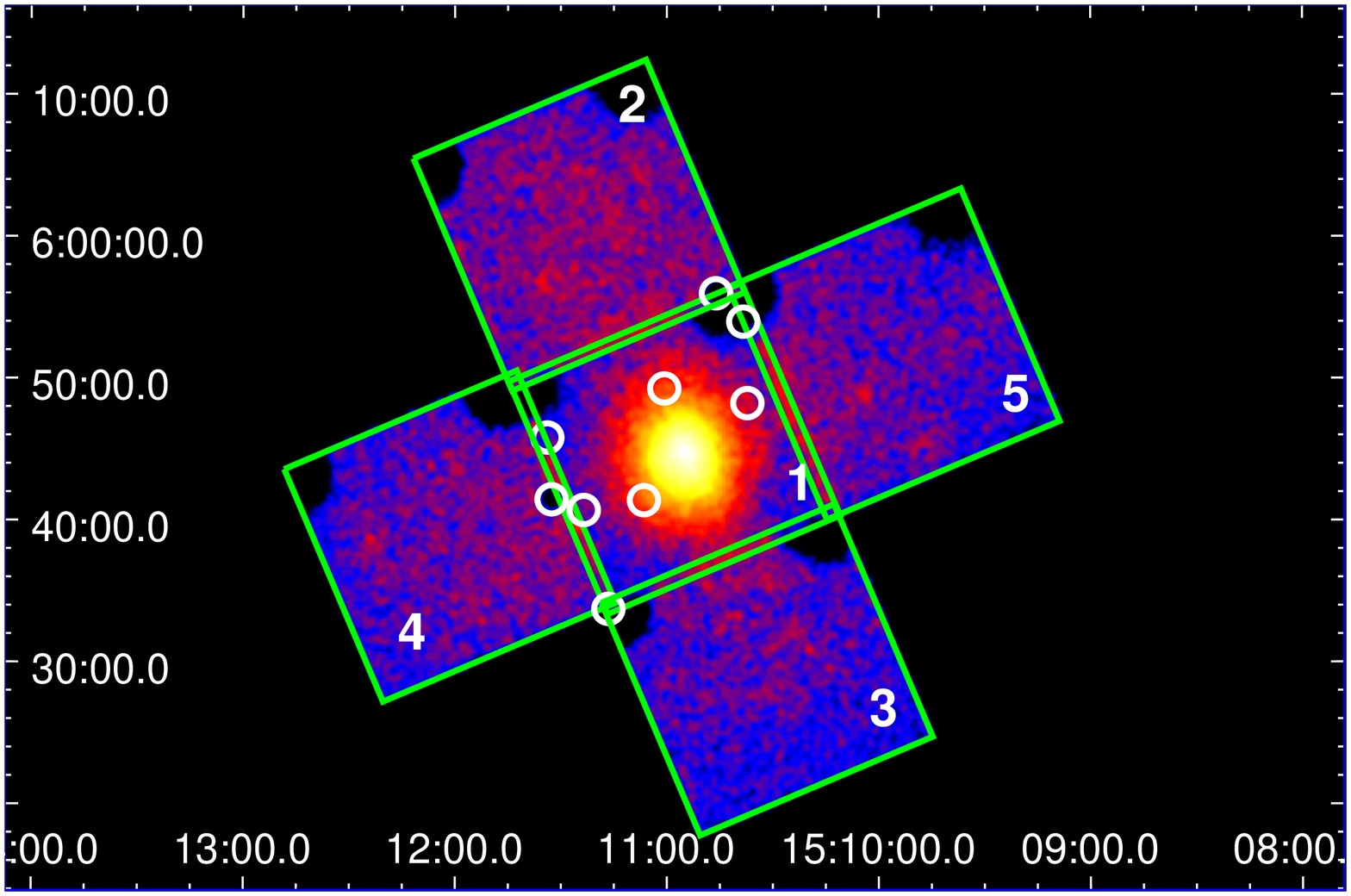}}}
\rotatebox{0}{\scalebox{0.25}{\includegraphics{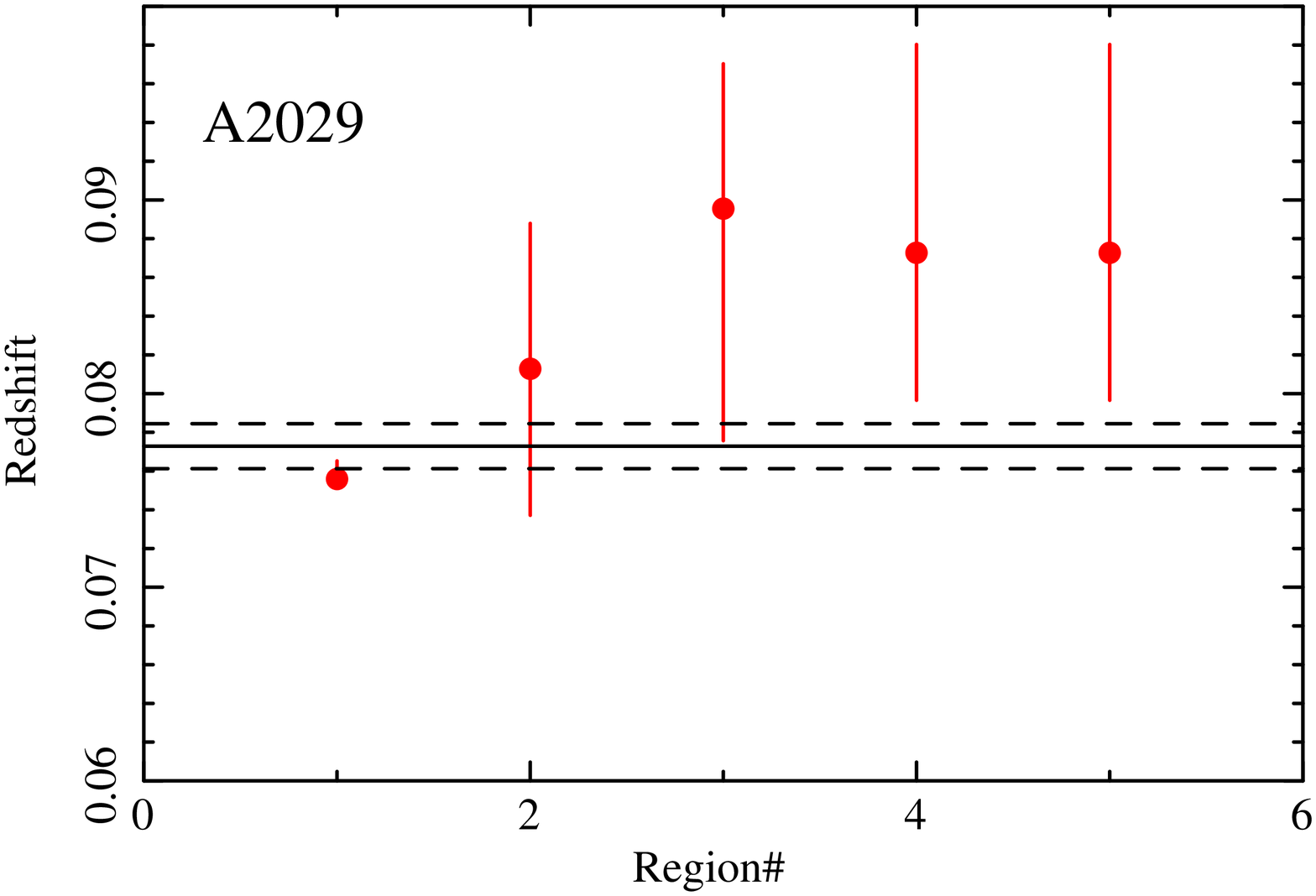}}}
\end{center}
\caption{(left) XIS-1 images in the 0.5--10~keV band, and (right)
  results of (i) the large-scale redshift measurements. From top to
  bottom, Centaurus, A2199, A3667, A133, and A2029 are displayed. In
  the left panels, the XIS images are smoothed by Gaussian
  $\sigma=16\arcsec$ but were not corrected for exposure and telescope
  vignetting. The two corners of the CCD chip illuminated by the
  calibration sources are excluded from the image. The spectral
  integration regions (i.e., the full XIS field of views of pointed
  observations) are shown by the green boxes and the locations of the
  point sources excluded from the spectral regions are marked with
  white circles. In the right panels, the red circles and the error
  bars indicate the best-fit redshift values and $1\sigma$ statistical
  uncertainties. The horizontal solid lines show the optical redshift
  of the cluster, and the dashed lines indicate the interval of
  $1\sigma$ systematic error ($\pm 0.1$\%). }\label{fig1}
\end{figure*}

For (i) the large-scale velocity measurements, we defined the spectral
regions by the size of the XIS FOV. For five clusters (Centaurus,
A2199, A3667, A133, and A2029), the spectra were extracted from each
FOV of multiple pointing data as shown in Fig.~\ref{fig1}. For the
remaining clusters (Perseus, A2255, and A2142), only the central
pointing data were used. Note that for the north region of A3667,
spectra from a common area of two datasets (A3667 NW1 and NW2) were
analyzed and are referred to as A3667 NW.  For the multiple pointing
data, the region identification numbers designated in Table~\ref{tab3}
are shown in Fig.~\ref{fig1}.

\begin{figure*}
\begin{center}
\rotatebox{0}{\scalebox{0.25}{\includegraphics{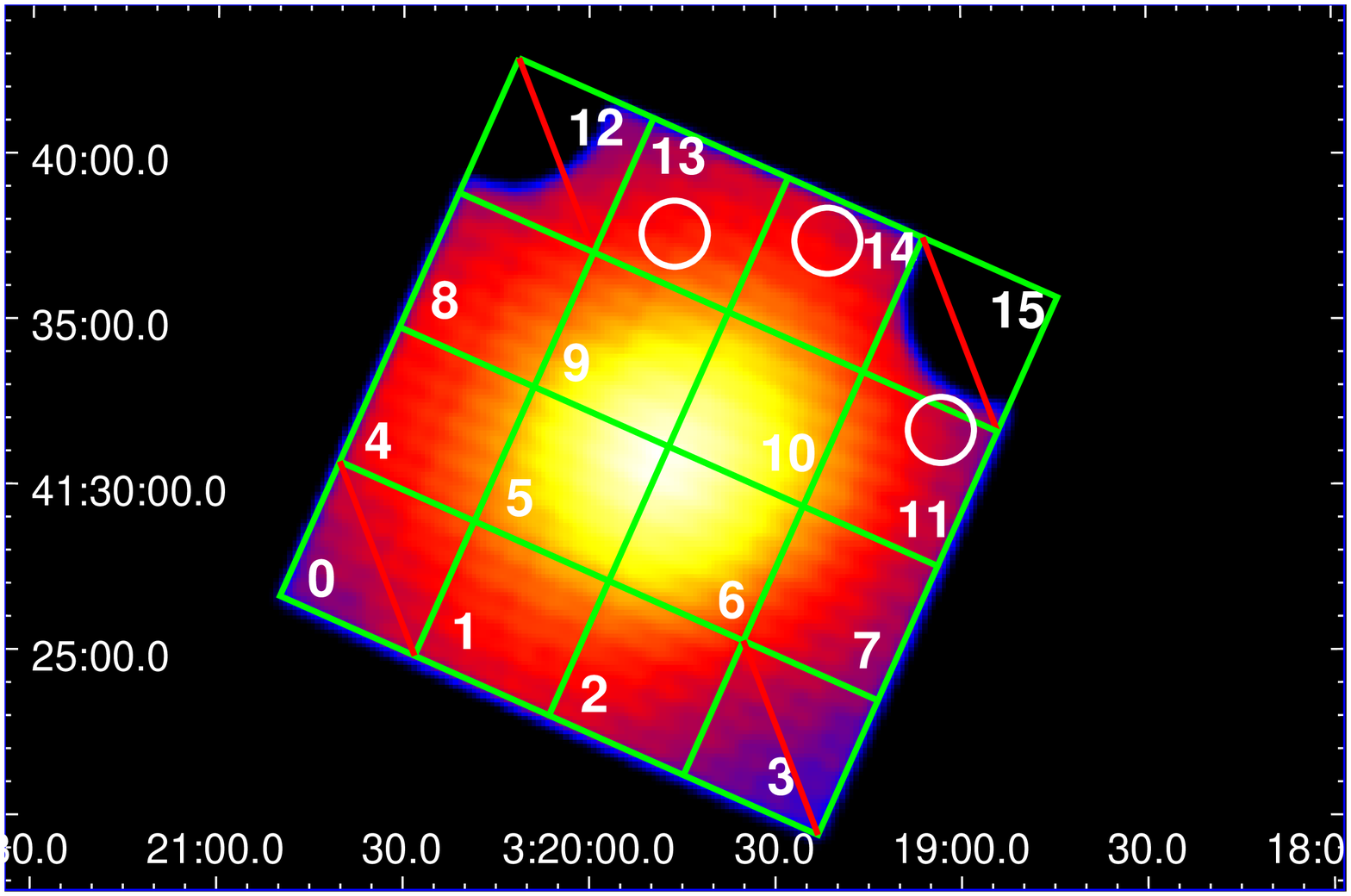}}}
\rotatebox{0}{\scalebox{0.26}{\includegraphics{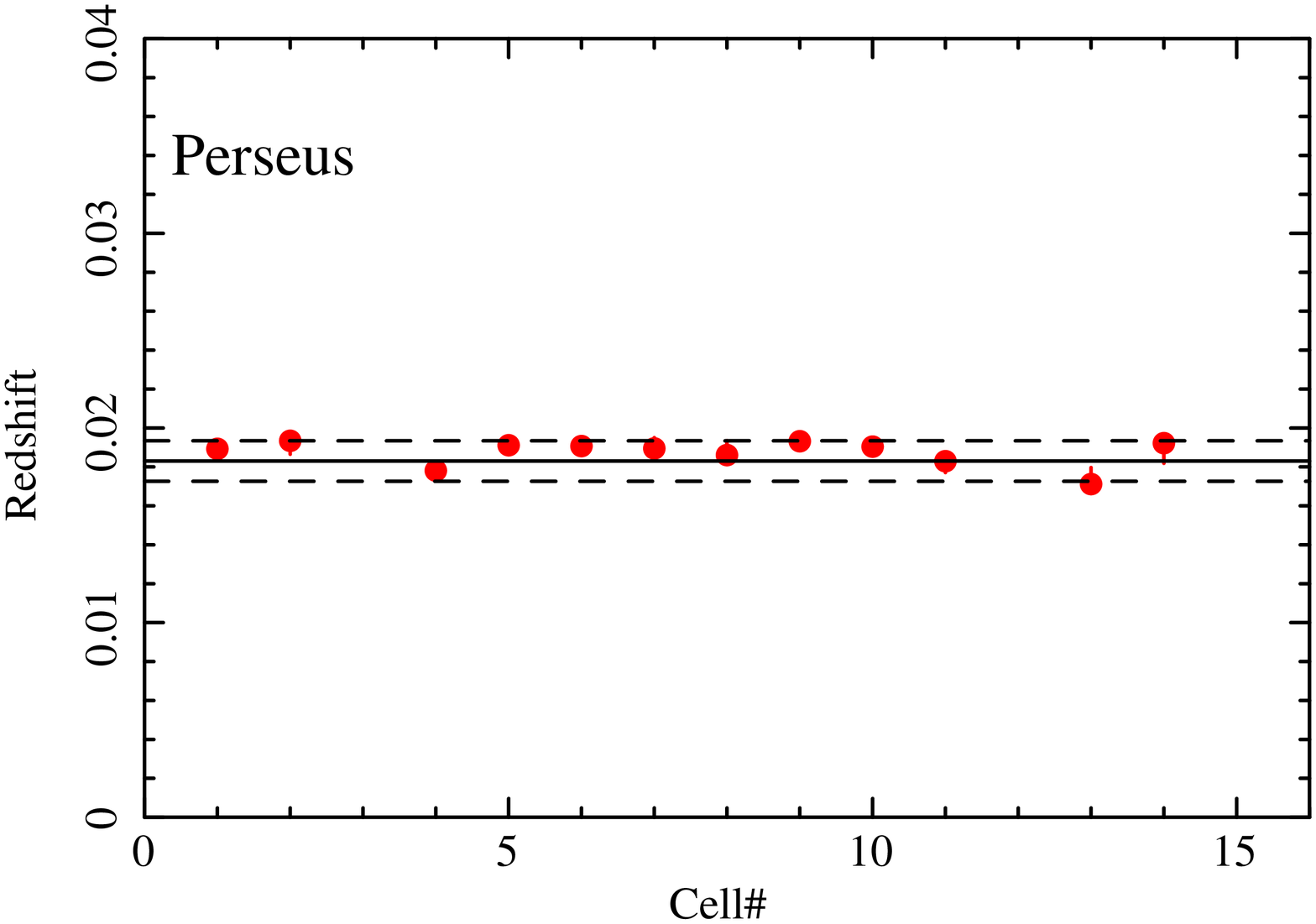}}}

\rotatebox{0}{\scalebox{0.25}{\includegraphics{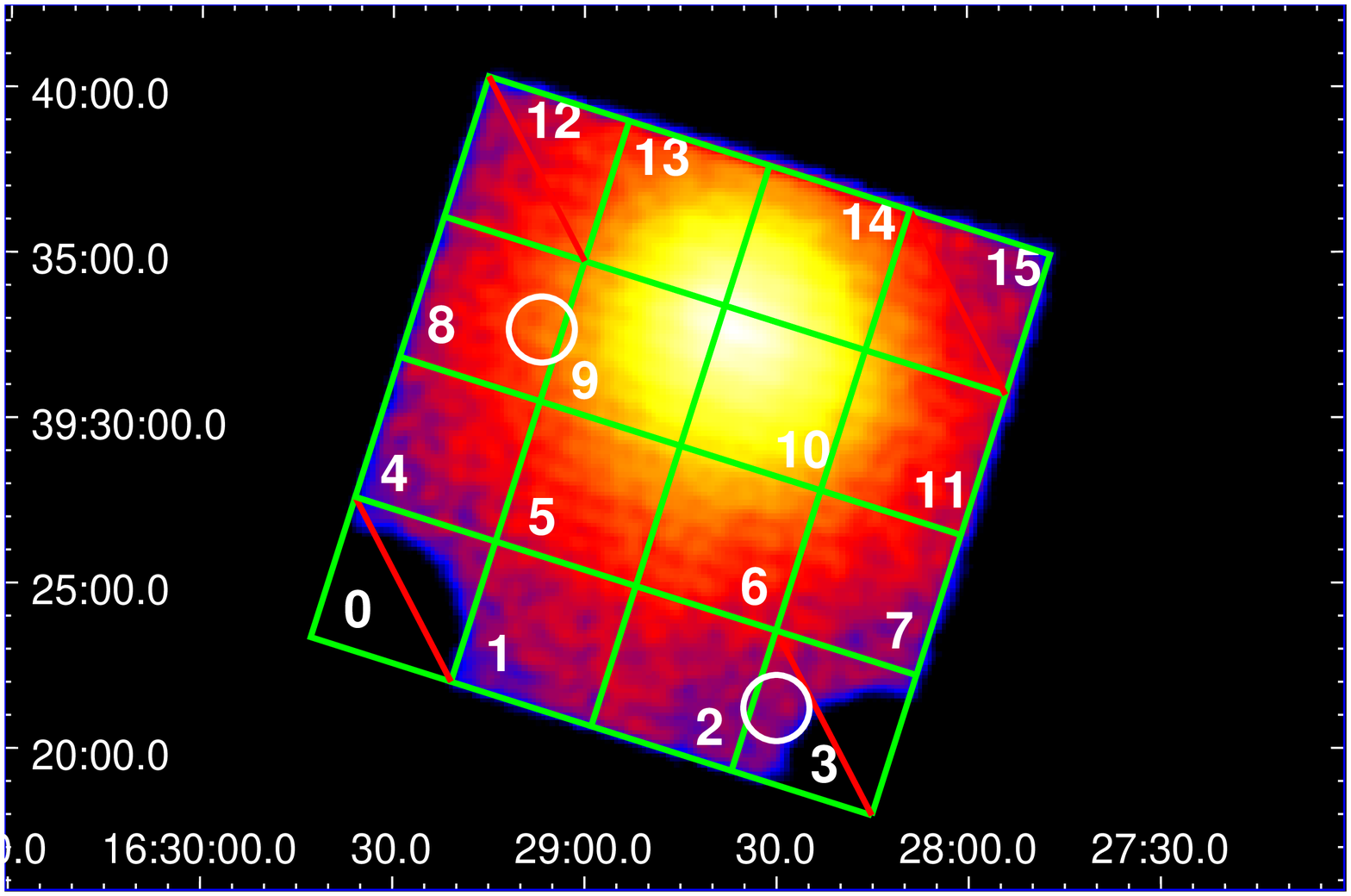}}}
\rotatebox{0}{\scalebox{0.26}{\includegraphics{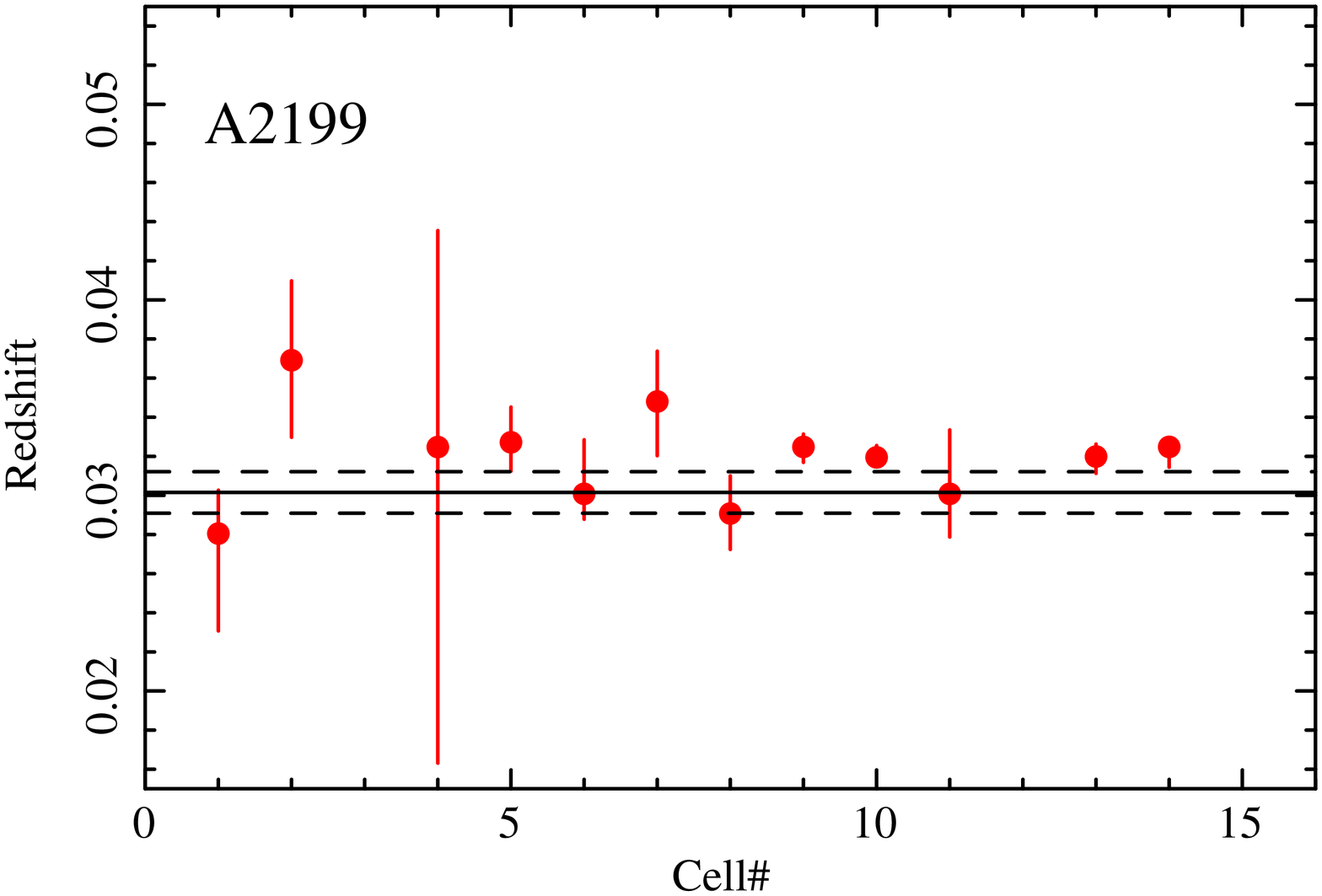}}}

\rotatebox{0}{\scalebox{0.25}{\includegraphics{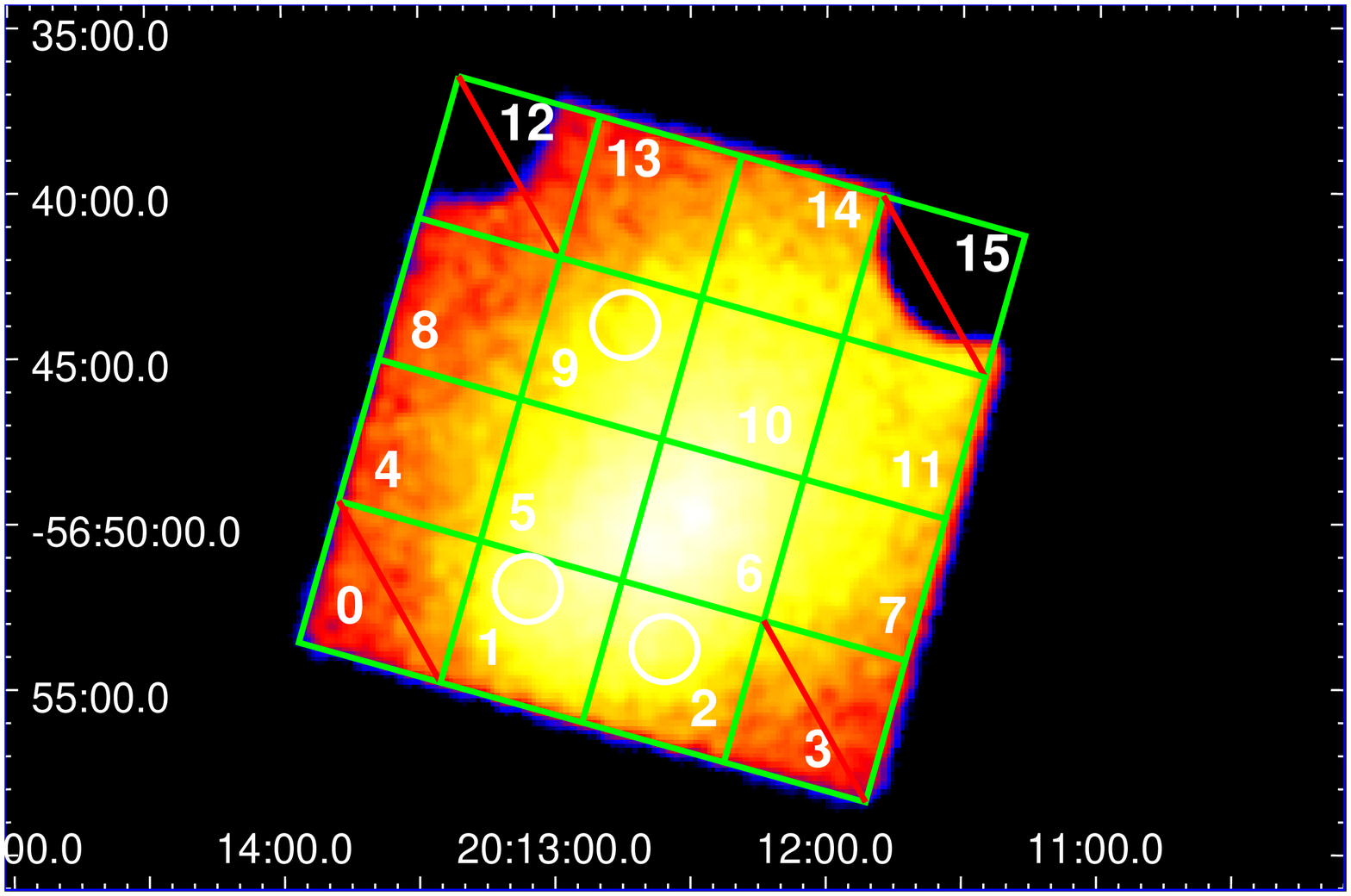}}}
\rotatebox{0}{\scalebox{0.26}{\includegraphics{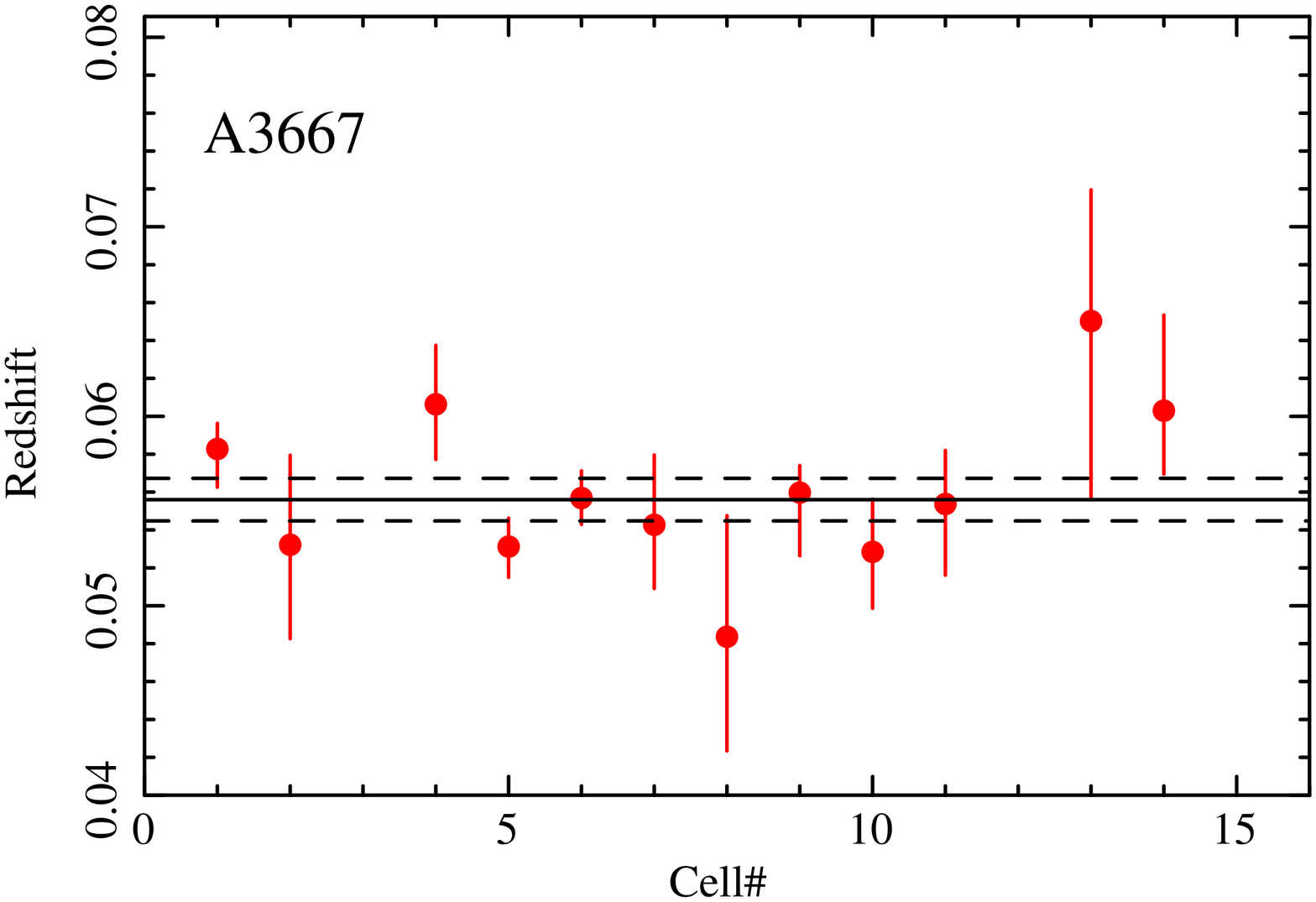}}}

\rotatebox{0}{\scalebox{0.25}{\includegraphics{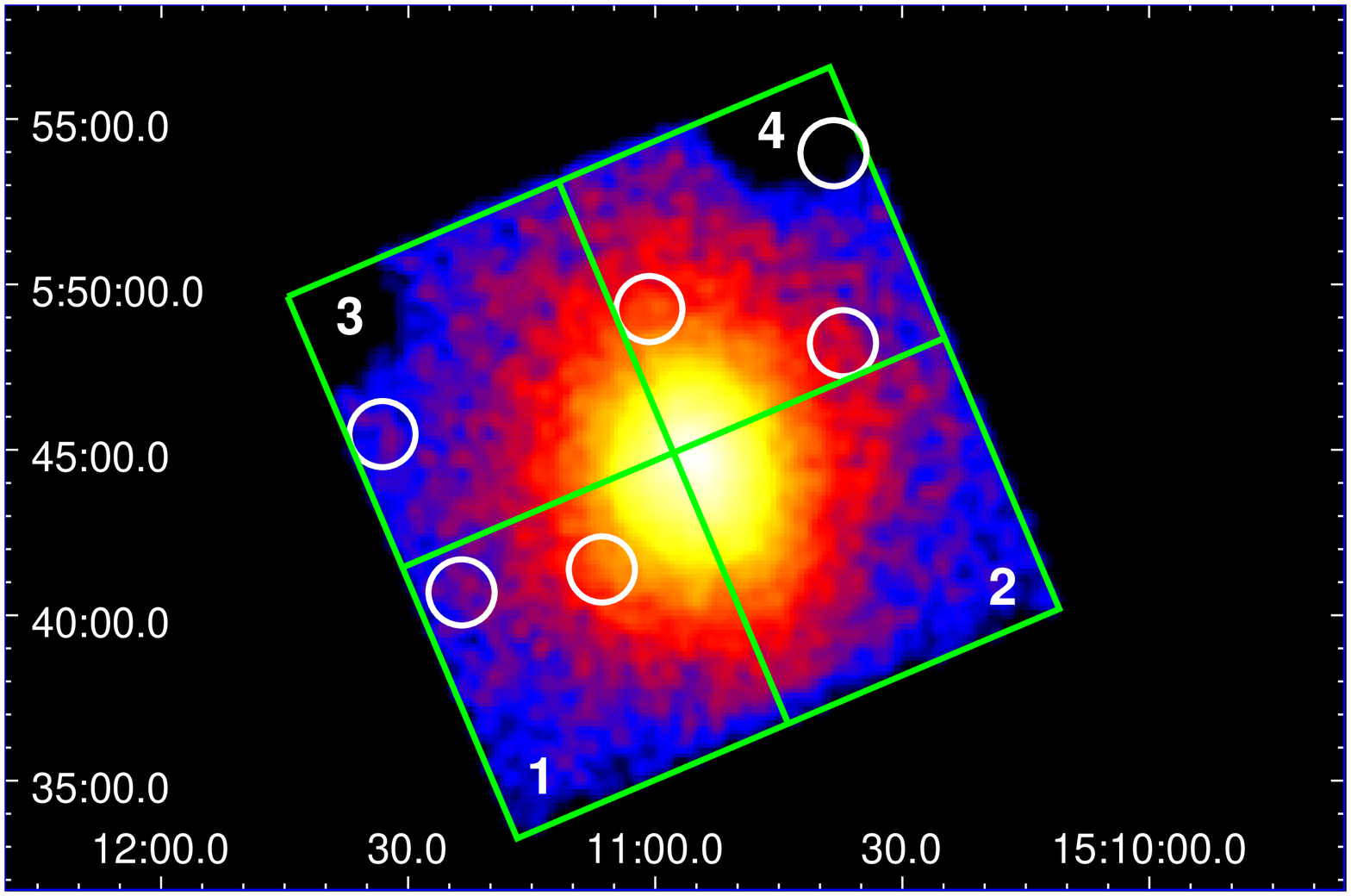}}}
\rotatebox{0}{\scalebox{0.26}{\includegraphics{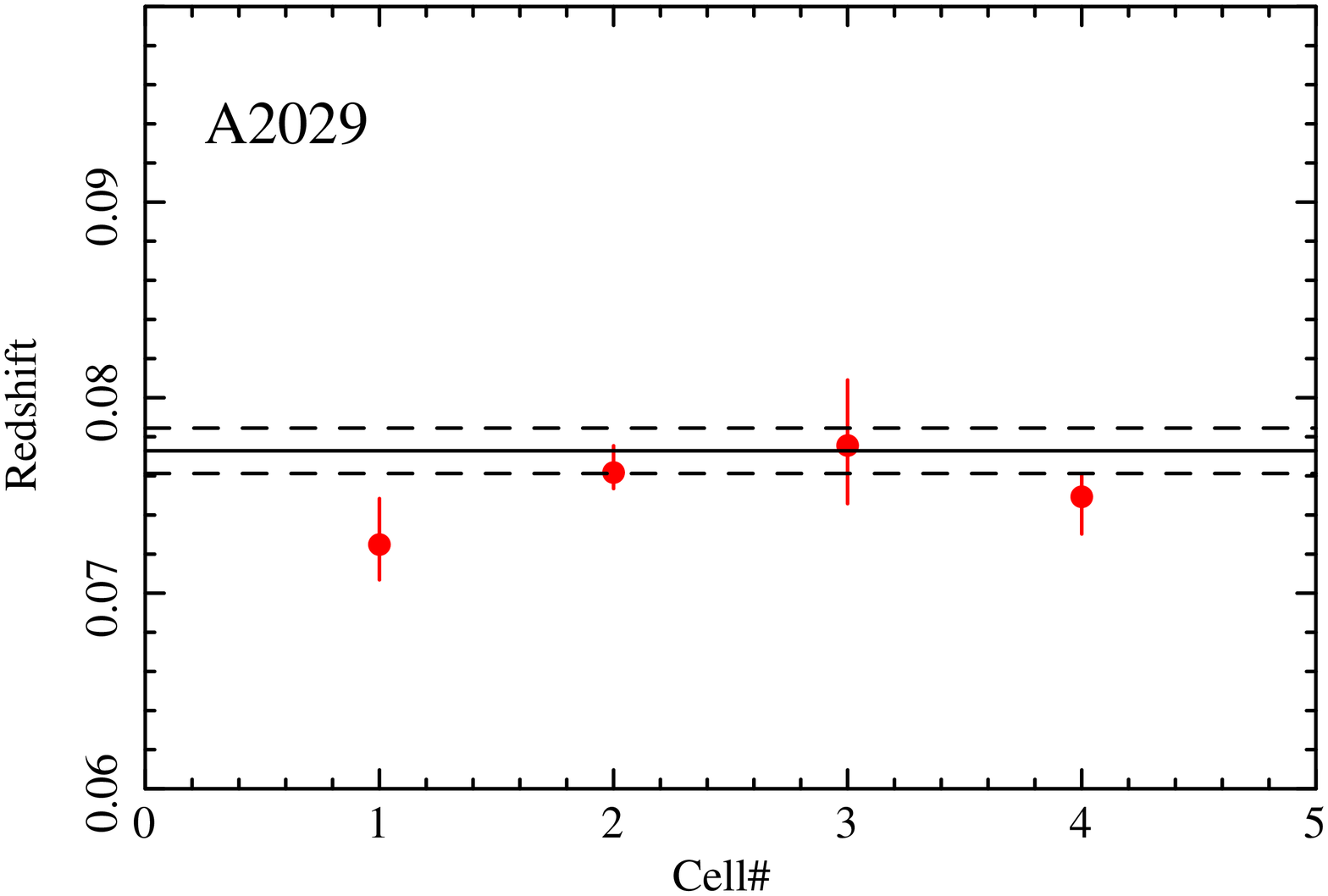}}}

\rotatebox{0}{\scalebox{0.25}{\includegraphics{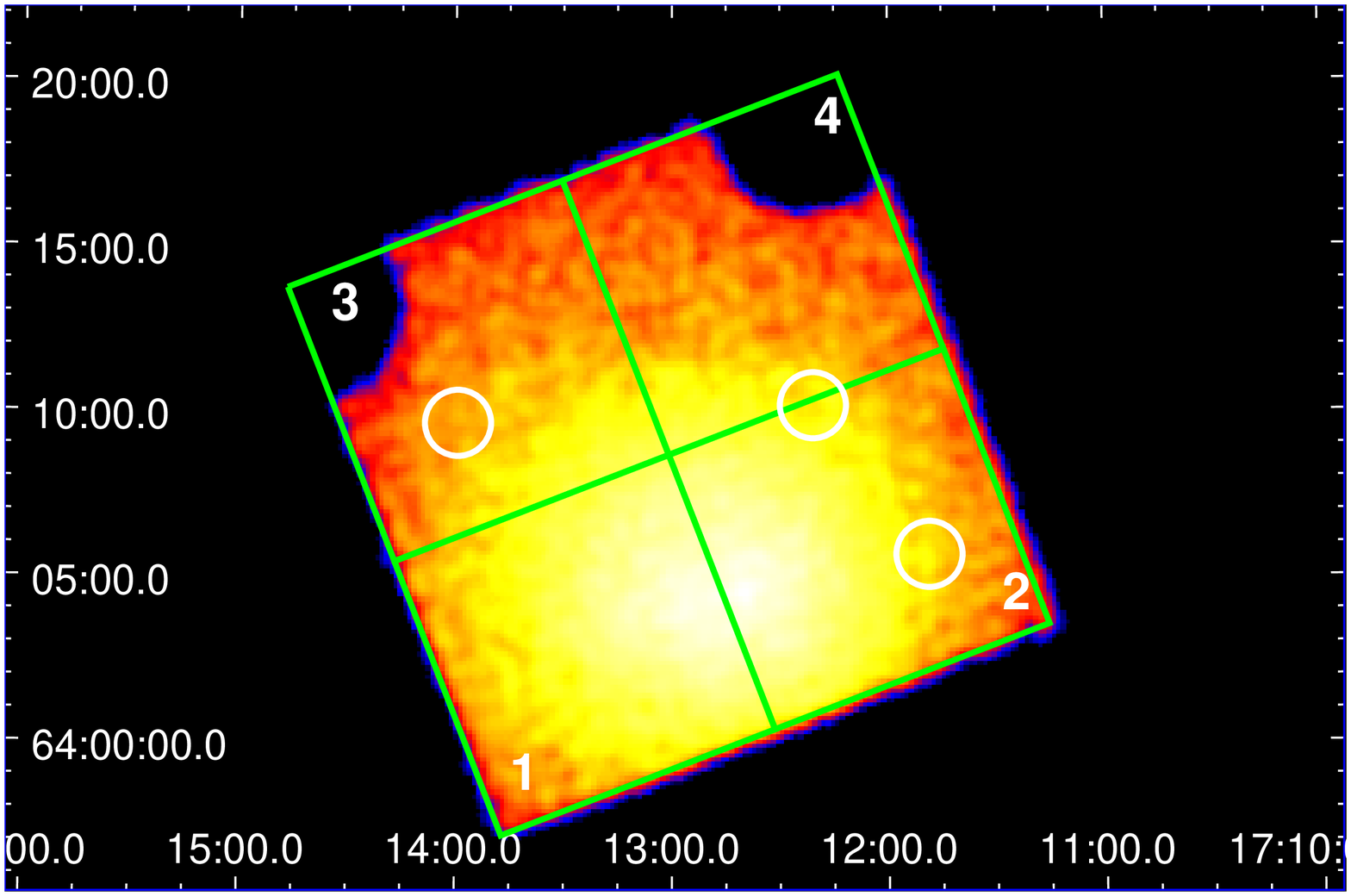}}}
\rotatebox{0}{\scalebox{0.26}{\includegraphics{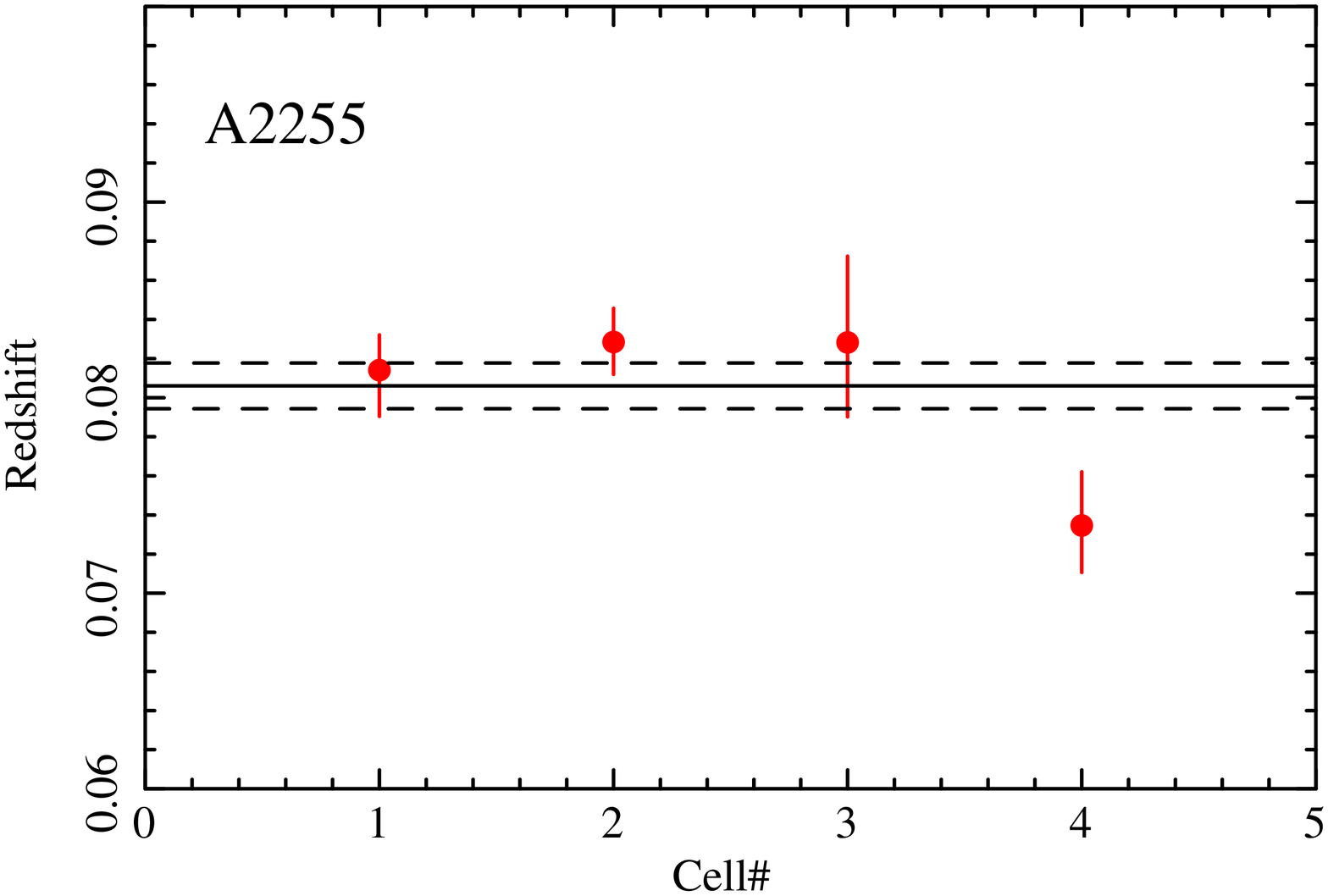}}}

\rotatebox{0}{\scalebox{0.25}{\includegraphics{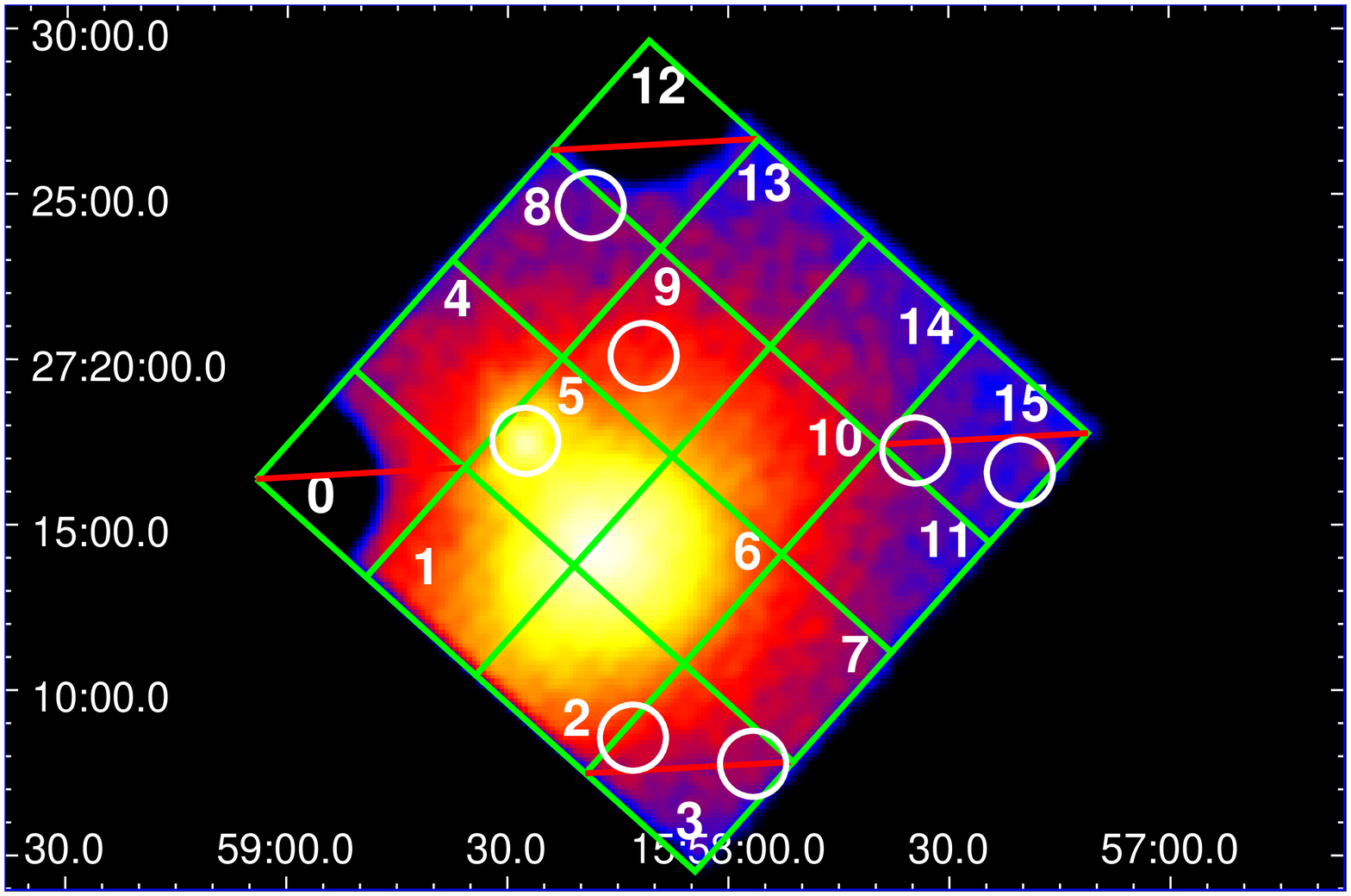}}}
\rotatebox{0}{\scalebox{0.26}{\includegraphics{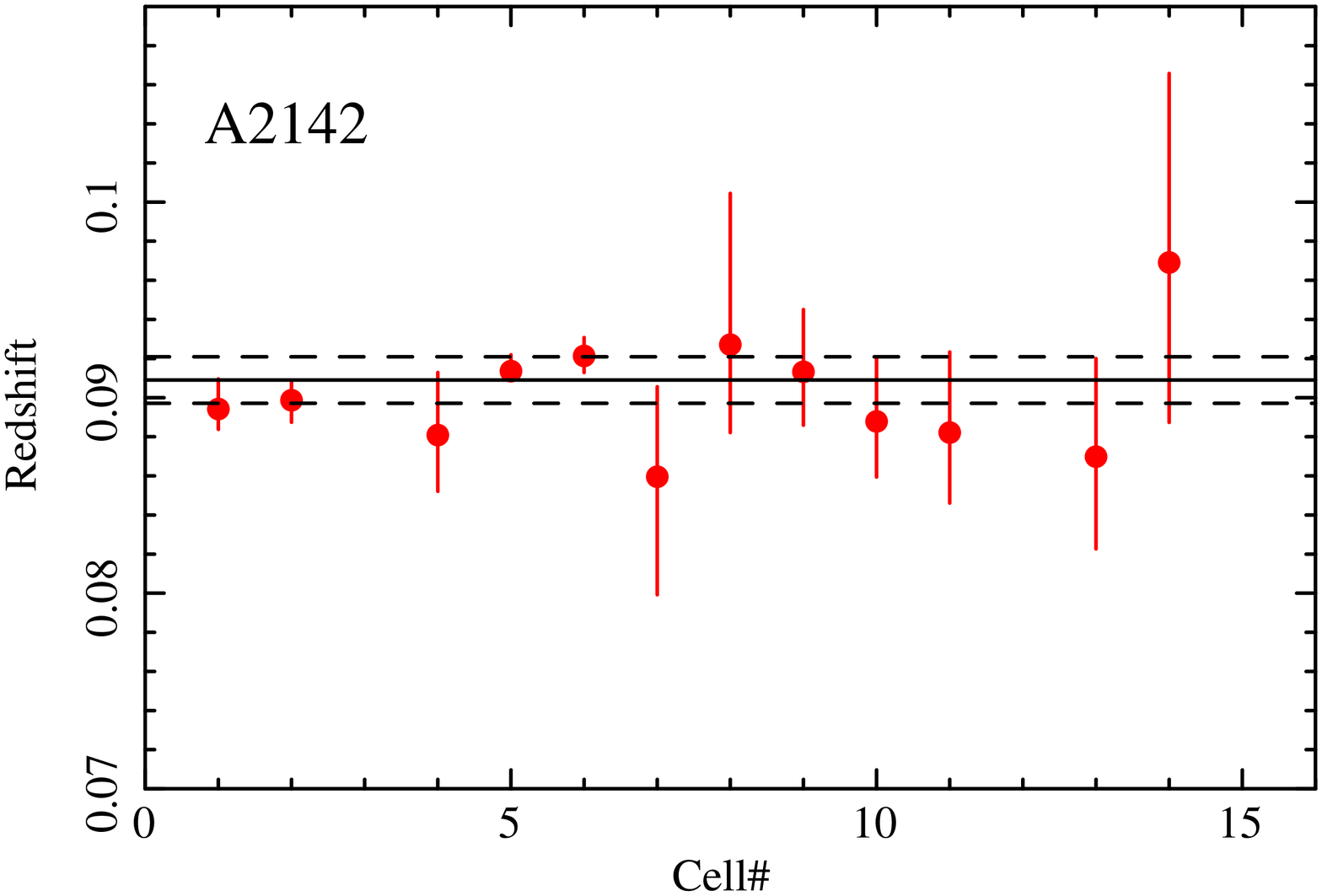}}}
\end{center}
\caption{(left) XIS-1 images in the 0.5--10~keV band and (right)
  results of (ii) the small-scale redshift measurements. From top to
  bottom, Perseus, A2199, A3667,
  A2029, A2255, and A2142 are shown. In the
  left panels, the cells used for spectral extraction are shown with
  the green boxes and the cell numbers are indicated. In the right
  panels, the meanings of the symbols are the same as those in
  Fig.~\ref{fig1}.}\label{fig2}
\end{figure*}

For (ii) the small-scale measurements, we divided the XIS FOVs into
smaller cells. The cell sizes were chosen so that the Fe-K line
spectrum contained at least 100 photons except for some very faint
areas near the edge of the XIS cameras.  This allows a redshift
determination to the accuracy better than 10\% in majority of cells.
As shown in Fig.~\ref{fig2}, XIS spectra were extracted from
$4\times4$ cells (i.e., 1 cell = $4\arcmin.5\times4\arcmin.5$) in
Perseus, A2199, A3667, and A2142, and $2\times2$ cells (i.e., 1 cell =
$9\arcmin\times9\arcmin$) in A2029 and A2255.  Note that we show
fitting results for all the cells in the table and figures, but
exclude the faint areas near the CCD edge for which only loose constraint is obtained when we estimate the upper limits on bulk velocity.

\subsection{Redshift determination by spectral fitting}\label{subsec:fitting}
\begin{figure}
\begin{center}
\rotatebox{0}{\scalebox{0.3}{\includegraphics{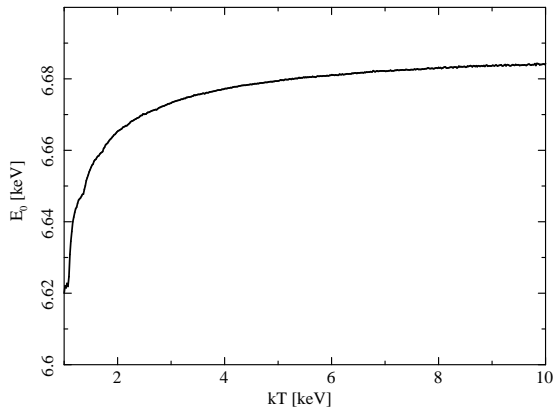}}}
\end{center}
\caption{Centroid energy of the He-like Fe-K line complex in the rest frame
  with $E_0$~(keV) as a function of gas temperature, $kT$~(keV). $E_0$
  was calculated by assuming the APEC model and the XIS energy
  resolution of 150~eV FWHM. }\label{fig3}
\end{figure}

The redshift of the Fe-K line was calculated from the relation $z
=(E_0-E_{\rm obs})/E_{\rm obs}$, where $E_0$ and $E_{\rm obs}$ are the
centroid energies of the He-like Fe-K line complex in the rest
flame and the observed flame, respectively. Fig.~\ref{fig3} shows the
centroid energy $E_0$ expected in the case such that the He-like
Fe-K line complex is convolved with the XIS energy resolution at a
full width at half maximum (FWHM) of 150~eV at 6.7~keV. To be strict,
because $E_0$ changes with gas temperature, both $E_0$ and $E_{\rm
  obs}$ are required to determine $z$ in each cluster region.

To quantify the rest-frame line energy $E_0$, the mean gas temperature
was determined by simultaneously fitting the XIS FI and BI spectra of
the central pointing of each cluster in the 0.7--10~keV band to
the Astrophysical Plasma Emission Code (APEC) model
\citep{2001ApJ...556L..91S}.  Note that for A133, the spectra taken
from A133 W was used and for the Centaurus cluster, the Cen45
subcluster was also analyzed.  The temperature, metal abundance, and
the normalization factor were allowed to vary, whereas the Galactic
hydrogen column density, $N_{\rm H}$, was fixed to a value from the
Leiden/Argentine/Bonn survey \citep{2005A&A...440..775K}.  The results
of the APEC model fitting and $E_0$ are shown in
Table~\ref{tab2}. Because the single-component APEC model did not give
a sufficient fit to the spectra of Centaurus (e.g.,
\cite{1999ApJ...525...58I,2011PASJ...63S.979S}) and Perseus (e.g.,
\cite{2009ApJ...705L..62T}), two-component APEC models were applied to
these two clusters. The parameters of the hotter component are shown
in Table~\ref{tab2}.

\begin{table*}
\tbl{Results of APEC model fitting to the XIS spectra}{ 
\begin{tabular}{llllllll} \hline 
Pointing & $N_{\rm H}$$^{\mathrm{a}}$& $kT$~(keV) & $Z$~(solar)$^{\mathrm{b}}$ & $z~(10^{-2})$ & $F_{\rm X}$$^{\mathrm{c}}$& $\chi^{2}$/d.o.f. & $E_0$~(keV)$^{\mathrm{d}}$ \\  \hline
Centaurus C	&	8.56	&	$	3.85 	\pm	0.02 	$	&	$	0.417 	\pm	0.007 	$	&	$	0.98 	\pm	0.02 	$	&	7.48 	&	$	5321 	/	5196	$	&	6.677 	\\
Cen45	&	8.56	&	$	4.12 	\pm	0.02 	$	&	$	0.315 	\pm	0.008 	$	&	$	1.17 	\pm	0.06 	$	&	2.36 	&	$	3760 	/	3721	$	&	6.677 	\\
Perseus	&	13.9	&	$	5.68 	\pm	0.01 	$	&	$	0.319 	\pm	0.003 	$	&	$	1.913 	\pm	0.003 	$	&	51.1 	&	$	9216 	/	8385	$	&	6.681 	\\
A2199 C	&	0.89	&	$	4.06 	\pm	0.01 	$	&	$	0.347 	\pm	0.006 	$	&	$	3.31 	\pm	0.01 	$	&	6.24 	&	$	5446 	/	5317	$	&	6.677 	\\
A3667 C	&	4.46	&	$	6.73 	\pm	0.05 	$	&	$	0.278 	\pm	0.010 	$	&	$	5.54 	\pm	0.06 	$	&	5.60 	&	$	4774 	/	4751	$	&	6.682 	\\
A133 W	&	1.59	&	$	3.27 	\pm	0.12 	$	&	$	0.313 	\pm	0.050 	$	&	$	5.70 	\pm	0.51 	$	&	0.11 	&	$	365 	/	305	$	&	6.675 	\\
A2029 C	&	3.26	&	$	7.50 	\pm	0.09 	$	&	$	0.388 	\pm	0.016 	$	&	$	7.60 	\pm	0.08 	$	&	8.56 	&	$	2732 	/	2728	$	&	6.683 	\\
A2255 C	&	2.59	&	$	5.88 	\pm	0.07 	$	&	$	0.182 	\pm	0.012 	$	&	$	8.04 	\pm	0.06 	$	&	1.46 	&	$	2821 	/	2863	$	&	6.681 	\\
A2142	&	3.78	&	$	8.04 	\pm	0.07 	$	&	$	0.235 	\pm	0.010 	$	&	$	9.03 	\pm	0.05 	$	&	6.50 	&	$	5574 	/	5504	$	&	6.683 	\\\hline
\end{tabular}}\label{tab2}
\begin{tabnote}
$^{\mathrm{a}}$ The Galactic hydrogen column density ($10^{20}~{\rm cm^{-2}}$).
$^{\mathrm{b}}$ The metal abundance of the ICM (solar). The abundance table in \citet{1989GeCoA..53..197A} was assumed. 
$^{\mathrm{c}}$ The X-ray flux in the 2--10~keV band ($10^{-11}~{\rm erg\,s^{-1}cm^{-2}}$) for each XIS pointing. 
$^{\mathrm{d}}$ The centroid energy of the He-like Fe-K line complex in the rest frame (\S\ref{subsec:fitting}).
\end{tabnote}
\end{table*}

\begin{figure}
\begin{center}
\rotatebox{0}{\scalebox{0.3}{\includegraphics{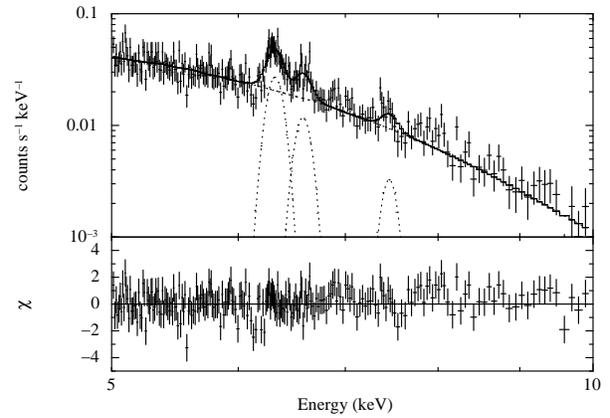}}}
\end{center}
\caption{Example of Gaussian fitting to the Fe-K line spectra. In the
  upper panel, the crosses show the XIS spectrum observed in A3667
  cell No.~6. The solid line represents the best-fit model and the
  three Gaussian lines are indicated with dashed lines. In the bottom
  panel, the residuals of fit are shown.}\label{fig4}
\end{figure}

To accurately measure the line centroid $E_{\rm obs}$ for the cases
(i) and (ii), the XIS spectra were analyzed in the following manner,
which is the same as that reported by \citet{2007PASJ...59S.351O}. The
5--10~keV spectra were fitted to a model consisting of continuum
emission represented by the APEC model with a metallicity of
$Z=0$~solar and three Gaussian lines for He-like Fe K line complex, H-like
Fe K$\alpha$, and a blend of He-like Ni K$\alpha$ and He-like Fe
K$\beta$ lines. The gas temperature of the APEC model and $E_0$ were
fixed to the mean values derived above. The Gaussian width was assumed
to be 0 because the intrinsic line width is smaller than the
instrumental energy resolution and it does not affect the present
analysis. Figure~\ref{fig4} shows an example of the spectral fitting.
For A2255 cell No.~4, spectra from a common area of two data sets were
co-added because they gave statistically consistent fitting results:
$z=0.0720 \pm 0.0037$ (A2255 C) and $z=0.0743 \pm 0.0026 $ (A2255 NW).

The XIS spectral bins in each cluster region were grouped so that each
spectral bin contains more than 20 photons and the chi-squared fitting
was performed by {\tt XSPEC} version 12.8.0. The right panels in
Figures \ref{fig1} and \ref{fig2} show the ICM redshift values
obtained from the spectral fitting for (i) and (ii), respectively.

\subsection{Systematic error of the Fe line centroid energy}\label{subsec:accuracy}
To estimate the systematic error of the iron line energy, we checked
the centroid energies of the Mn-K lines from the build-in calibration
sources that illuminates two corners of each XIS chip. We extracted
spectra of the calibration sources and fitted the 4--8~keV spectra
with two Gaussians for the Mn K$\alpha$ and K$\beta$ along with the
bremsstrahlung continuum component. The resultant line centroid of Mn
K$\alpha$ line agreed with the expected value of 5.894~keV within $\pm
0.1$\%. This result is consistent with that reported by
\citet{2009PASJ...61S...1O}, who performed a precise in-orbit
calibration of the XIS energy-scale by using the build-in calibration
sources and 1E0102.2--7219. We thus assigned $\pm0.1$\% to the
$1\sigma$ systematic error on the XIS energy scale and calculated the
$1\sigma$ systematic uncertainty of the redshift ($\sigma_{z, {\rm
    sys}}$) and line-of-sight velocity ($\sigma_{v, {\rm sys}}$). The
values for the sample clusters, shown in
Tables~\ref{tab3}--\ref{tab4}, were typically $\sigma_{z, {\rm sys}} =
0.001$ and $\sigma_{v, {\rm sys}} = 300~{\rm km\,s^{-1}}$ for
$z=0.01$.

\subsection{Constraints on the gas bulk velocity}\label{subsec:bulk_constraint}
Figure~\ref{fig1} shows that the ICM redshifts obtained in the case of
(i) the large-scale measurements. For the eight clusters the fitted
redshifts agree with the optical values ($z_{\rm cl}$) within the
systematic errors.  As shown in Fig.~\ref{fig2}, in case of (ii) the
small-scale analyses, the redshifts in Perseus and A2142 were
consistent with $z_{\rm cl}$ within the systematic errors. On the
contrary, some cells in A2199, A3667, A2029, and A2255, showed a shift
of $\Delta z = (z - z_{\rm cl})$ in excess of the $1\sigma$ systematic
error range.

The line-of-sight bulk velocity (i.e., the radial velocity relative to
the optical cluster redshift) in each spectral region was calculated
by $\Delta v = c\Delta z$ and shown in Tables~\ref{tab3} and
\ref{tab4}. The above results indicate that significant bulk motions
exist in some cells in the cluster center. If both statistical and
systematic errors are considered, the significance of the bulk
velocity is $1.6-1.8\sigma$ for A2199 cell Nos.~2, 7, 9, and 14,
$1.6\sigma$ for A3667 cell No.~4, $2.1\sigma$ for A2029 cell No.~1,
and $2.6\sigma$ for A2255 cell No.~4, respectively.  Here the standard
deviation $\sigma$ is given by a quadrature sum of the $1\sigma$
statistical and systematic errors, $\sigma_{\rm tot} = (\sigma_{z,
  {\rm stat}}^2 + \sigma_{z, {\rm sys}}^2)^{1/2}$.

\begin{table*}
\tbl{Results of (i) the large-scale redshift and velocity measurements}{
\begin{tabular}{llllllllllll} \hline 
Pointing & Reg$^{\mathrm{a}}$ &$z_{\rm cl}$	&	$z$	&	$\sigma_{z, {\rm  stat}}$	&	$\sigma_{z, {\rm sys}}$	&	$\Delta z/\sigma_{z, {\rm tot}}$	$^{\mathrm{b}}$
   &	$\Delta v$	&	$\sigma_{v, {\rm stat}}$	&	$\sigma_{v, {\rm sys}}$	& $|\Delta v|$$^{\mathrm{c}}$	& $\chi^2$/d.o.f.\\ 
   & & ($10^{-2}$) & ($10^{-2}$) & ($10^{-2}$) & ($10^{-2}$) & & (${\rm km\,s^{-1}}$) & (${\rm km\,s^{-1}}$) & (${\rm km\,s^{-1}}$) & (${\rm km\,s^{-1}}$) & \\\hline
Centaurus C	&	1	&	1.04	&	0.975 	&	0.007 	&	0.10 	&	$	-0.64 	$	&	$	-196 	$	&	20 	&	303 	&	$	<	500 	$	&	1083 	/	1037	\\	
Centaurus S	&	2	&	1.04	&	0.982 	&	0.021 	&	0.10 	&	$	-0.56 	$	&	$	-175 	$	&	62 	&	303 	&	$	<	484 	$	&	1052 	/	986	\\	
Centaurus N	&	3	&	1.04	&	1.022 	&	0.020 	&	0.10 	&	$	-0.18 	$	&	$	-54 	$	&	59 	&	303 	&	$	<	363 	$	&	1061 	/	972	\\	
Cen45	&	4	&	1.04	&	1.116 	&	0.045 	&	0.10 	&	$	+0.68 	$	&	$	+227 	$	&	134 	&	303 	&	$	<	558 	$	&	812 	/	832	\\	\hline
Perseus	&	-	&	1.83	&	1.931 	&	0.006 	&	0.10 	&	$	+0.99 	$	&	$	+304 	$	&	19 	&	305 	&	$	<	610 	$	&	2046 	/	1363	\\	\hline
A2199 C	&	1	&	3.02	&	3.20 	&	0.03 	&	0.10 	&	$	+1.67 	$	&	$	+535 	$	&	86 	&	309 	&	$	<	856 	$	&	758 	/	774	\\	
A2199 SE	&	2	&	3.02	&	3.17 	&	0.05 	&	0.10 	&	$	+1.28 	$	&	$	+444 	$	&	157 	&	309 	&	$	<	791 	$	&	505 	/	498	\\	
A2199 SW	&	3	&	3.02	&	3.17 	&	0.12 	&	0.10 	&	$	+0.97 	$	&	$	+450 	$	&	348 	&	309 	&	$	<	916 	$	&	344 	/	331	\\	
A2199 NW	&	4	&	3.02	&	2.67 	&	0.42 	&	0.10 	&	$	-0.80 	$	&	$	-1042 	$	&	1257 	&	309 	&	$	<	2336 	$	&	224 	/	232	\\	
A2199 NE	&	5	&	3.02	&	3.17 	&	0.16 	&	0.10 	&	$	+0.80 	$	&	$	+452 	$	&	474 	&	309 	&	$	<	1018 	$	&	250 	/	251	\\	\hline
A3667 NW	&	1	&	5.56	&	5.70 	&	0.24 	&	0.11 	&	$	+0.52 	$	&	$	+405 	$	&	711 	&	317 	&	$	<	1184 	$	&	337 	/	320	\\	
A3667 C	&	2	&	5.56	&	5.30 	&	0.31 	&	0.11 	&	$	-0.80 	$	&	$	-791 	$	&	936 	&	317 	&	$	<	1778 	$	&	210 	/	185	\\	
A3667 SE	&	3	&	5.56	&	5.66 	&	0.30 	&	0.11 	&	$	+0.33 	$	&	$	+309 	$	&	894 	&	317 	&	$	<	1257 	$	&	273 	/	281	\\	\hline
A133 W	&	1	&	5.66	&	5.85 	&	0.28 	&	0.11 	&	$	+0.64 	$	&	$	+577 	$	&	839 	&	317 	&	$	<	1474 	$	&	387 	/	354	\\	
A133 N	&	2	&	5.66	&	5.88 	&	0.42 	&	0.11 	&	$	+0.51 	$	&	$	+672 	$	&	1273 	&	317 	&	$	<	1983 	$	&	336 	/	370	\\	
A133 E	&	3	&	5.66	&	5.53 	&	0.45 	&	0.11 	&	$	-0.29 	$	&	$	-404 	$	&	1336 	&	317 	&	$	<	1777 	$	&	362 	/	354	\\	
A133 S	&	4	&	5.66	&	5.74 	&	0.64 	&	0.11 	&	$	+0.12 	$	&	$	+231 	$	&	1910 	&	317 	&	$	<	2167 	$	&	411 	/	397	\\	\hline
A2029 C	&	1	&	7.73	&	7.56 	&	0.06 	&	0.11 	&	$	-1.37 	$	&	$	-515 	$	&	189 	&	323 	&	$	<	889 	$	&	368 	/	342	\\	
A2029 N	&	2	&	7.73	&	8.13 	&	0.75 	&	0.11 	&	$	+0.52 	$	&	$	+1193 	$	&	2260 	&	323 	&	$	<	3476 	$	&	148 	/	139	\\	
A2029 S	&	3	&	7.73	&	8.95 	&	0.97 	&	0.11 	&	$	+1.25 	$	&	$	+3675 	$	&	2919 	&	323 	&	$	<	6612 	$	&	58 	/	58	\\	
A2029 E	&	4	&	7.73	&	8.73 	&	0.92 	&	0.11 	&	$	+1.08 	$	&	$	+2991 	$	&	2756 	&	323 	&	$	<	5766 	$	&	114 	/	117	\\	
A2029 W	&	5	&	7.73	&	9.83 	&	1.75 	&	0.11 	&	$	+1.20 	$	&	$	+6302 	$	&	5253 	&	323 	&	$	<	11565 	$	&	70 	/	70	\\	\hline
A2255 C	&	-	&	8.06	&	8.19 	&	0.10 	&	0.11 	&	$	+0.87 	$	&	$	+394 	$	&	315 	&	324 	&	$	<	846 	$	&	562 	/	569	\\	\hline
A2142	&	-	&	9.09	&	9.10 	&	0.05 	&	0.11 	&	$	+0.06 	$	&	$	+21 	$	&	142 	&	327 	&	$	<	378 	$	&	1134 	/	1105	\\	\hline
\end{tabular}}\label{tab3}
\begin{tabnote}
$^{\mathrm{a}}$ Region number. 
$^{\mathrm{b}}$ Significance of the observed Doppler shift of
  Fe-K line, calculated by the redshift difference relative to the
  optical value, $\Delta z = (z - z_{\rm cl})$, divided by the
  standard deviation $\sigma_{\rm tot}$ (see text).
$^{\mathrm{c}}$ The 68\% limit on the absolute velocity
  difference $|\Delta v|$. Both statistical and systematic errors are
  taken into account.
\end{tabnote}
\end{table*}

\begin{longtable}{llllllllllll}
\caption{Results of (ii) the small-scale redshift and velocity measurements}\label{tab4}
\hline 
Cluster & Cell	&	$z_{\rm cl}$	&	$z$	&	$\sigma_{z, {\rm  stat}}$	&	$\sigma_{z, {\rm sys}}$	&	$\Delta z/\sigma_{z, {\rm tot}}$
   &	$\Delta v$	&	$\sigma_{v, {\rm stat}}$	&	$\sigma_{v, {\rm sys}}$	& $|\Delta v|$	& $\chi^2$/d.o.f.\\ 
   &   & ($10^{-2}$) & ($10^{-2}$) & ($10^{-2}$) & ($10^{-2}$) & & (${\rm km\,s^{-1}}$) & (${\rm km\,s^{-1}}$) & (${\rm km\,s^{-1}}$) & (${\rm km\,s^{-1}}$)& \\
\hline   
\endfirsthead 
\hline
Cluster & Cell	&	$z_{\rm cl}$	&	$z$	&	$\sigma_{z, {\rm  stat}}$	&	$\sigma_{z, {\rm sys}}$	&	$\Delta z/\sigma_{z, {\rm tot}}$
   &	$\Delta v$	&	$\sigma_{v, {\rm stat}}$	&	$\sigma_{v, {\rm sys}}$	& $|\Delta v|$	& $\chi^2$/d.o.f.\\ 
   &   & ($10^{-2}$) & ($10^{-2}$) & ($10^{-2}$) & ($10^{-2}$) & & (${\rm km\,s^{-1}}$) & (${\rm km\,s^{-1}}$) & (${\rm km\,s^{-1}}$) & (${\rm km\,s^{-1}}$)& \\
   \hline
\endhead
\hline
\endfoot
  \hline
\endlastfoot
\hline
Perseus	&	1	&	1.83	&	1.89 	&	0.010 	&	0.10 	&	$	+0.61 	$	&	$	+188 	$	&	31 	&	305 	&	$	<	495 	$	&	600 	/	525	\\
	&	2	&	1.83	&	1.93 	&	0.051 	&	0.10 	&	$	+0.92 	$	&	$	+313 	$	&	152 	&	305 	&	$	<	654 	$	&	519 	/	456	\\
	&	4	&	1.83	&	1.78 	&	0.034 	&	0.10 	&	$	-0.46 	$	&	$	-148 	$	&	101 	&	305 	&	$	<	470 	$	&	526 	/	524	\\
	&	5	&	1.83	&	1.91 	&	0.008 	&	0.10 	&	$	+0.80 	$	&	$	+245 	$	&	25 	&	305 	&	$	<	551 	$	&	1245 	/	1056	\\
	&	6	&	1.83	&	1.91 	&	0.011 	&	0.10 	&	$	+0.75 	$	&	$	+230 	$	&	32 	&	305 	&	$	<	538 	$	&	1193 	/	1069	\\
	&	7	&	1.83	&	1.90 	&	0.048 	&	0.10 	&	$	+0.58 	$	&	$	+195 	$	&	145 	&	305 	&	$	<	534 	$	&	414 	/	451	\\
	&	8	&	1.83	&	1.86 	&	0.038 	&	0.10 	&	$	+0.28 	$	&	$	+93 	$	&	113 	&	305 	&	$	<	418 	$	&	382 	/	433	\\
	&	9	&	1.83	&	1.93 	&	0.011 	&	0.10 	&	$	+0.99 	$	&	$	+305 	$	&	34 	&	305 	&	$	<	613 	$	&	1071 	/	971	\\
	&	10	&	1.83	&	1.90 	&	0.021 	&	0.10 	&	$	+0.71 	$	&	$	+222 	$	&	62 	&	305 	&	$	<	534 	$	&	1095 	/	1021	\\
	&	11	&	1.83	&	1.83 	&	0.044 	&	0.10 	&	$	-0.01 	$	&	$	-2 	$	&	132 	&	305 	&	$	<	335 	$	&	457 	/	443	\\
	&	13	&	1.83	&	1.71 	&	0.067 	&	0.10 	&	$	-0.97 	$	&	$	-354 	$	&	200 	&	305 	&	$	<	719 	$	&	308 	/	299	\\
	&	14	&	1.83	&	1.92 	&	0.063 	&	0.10 	&	$	+0.77 	$	&	$	+276 	$	&	188 	&	305 	&	$	<	634 	$	&	433 	/	387	\\ \hline
A2199	&	1	&	3.02	&	2.80 	&	0.36 	&	0.10 	&	$	-0.57 	$	&	$	-646 	$	&	1080 	&	309 	&	$	<	1769 	$	&	9 	/	19	\\
	&	2	&	3.02	&	3.69 	&	0.40 	&	0.10 	&	$	+1.62 	$	&	$	+2014 	$	&	1201 	&	309 	&	$	<	3253 	$	&	15 	/	18	\\
	&	4	&	3.02	&	3.25 	&	1.36 	&	0.10 	&	$	+0.17 	$	&	$	+682 	$	&	4085 	&	309 	&	$	<	4779 	$	&	14 	/	26	\\
	&	5	&	3.02	&	3.27 	&	0.16 	&	0.10 	&	$	+1.30 	$	&	$	+755 	$	&	491 	&	309 	&	$	<	1335 	$	&	40 	/	49	\\
	&	6	&	3.02	&	3.01 	&	0.20 	&	0.10 	&	$	-0.05 	$	&	$	-34 	$	&	610 	&	309 	&	$	<	718 	$	&	42 	/	52	\\
	&	7	&	3.02	&	3.48 	&	0.27 	&	0.10 	&	$	+1.61 	$	&	$	+1381 	$	&	801 	&	309 	&	$	<	2239 	$	&	24 	/	23	\\
	&	8	&	3.02	&	2.91 	&	0.19 	&	0.10 	&	$	-0.52 	$	&	$	-336 	$	&	564 	&	309 	&	$	<	979 	$	&	25 	/	29	\\
	&	9	&	3.02	&	3.25 	&	0.07 	&	0.10 	&	$	+1.82 	$	&	$	+685 	$	&	216 	&	309 	&	$	<	1062 	$	&	203 	/	190	\\
	&	10	&	3.02	&	3.20 	&	0.05 	&	0.10 	&	$	+1.51 	$	&	$	+525 	$	&	158 	&	309 	&	$	<	873 	$	&	212 	/	224	\\
	&	11	&	3.02	&	3.01 	&	0.27 	&	0.10 	&	$	-0.04 	$	&	$	-35 	$	&	821 	&	309 	&	$	<	912 	$	&	45 	/	43	\\
	&	13	&	3.02	&	3.20 	&	0.07 	&	0.10 	&	$	+1.41 	$	&	$	+538 	$	&	224 	&	309 	&	$	<	919 	$	&	156 	/	153	\\
	&	14	&	3.02	&	3.25 	&	0.07 	&	0.10 	&	$	+1.86 	$	&	$	+686 	$	&	200 	&	309 	&	$	<	1054 	$	&	177 	/	144	\\ \hline
A3667	&	1	&	5.56	&	5.83 	&	0.17 	&	0.11 	&	$	+1.35 	$	&	$	+804 	$	&	503 	&	317 	&	$	<	1399 	$	&	73 	/	69	\\
	&	2	&	5.56	&	5.32 	&	0.48 	&	0.11 	&	$	-0.48 	$	&	$	-715 	$	&	1452 	&	317 	&	$	<	2201 	$	&	29 	/	36	\\
	&	4	&	5.56	&	6.06 	&	0.30 	&	0.11 	&	$	+1.58 	$	&	$	+1510 	$	&	903 	&	317 	&	$	<	2467 	$	&	28 	/	31	\\
	&	5	&	5.56	&	5.31 	&	0.16 	&	0.11 	&	$	-1.31 	$	&	$	-742 	$	&	469 	&	317 	&	$	<	1308 	$	&	100 	/	138	\\
	&	6	&	5.56	&	5.57 	&	0.14 	&	0.11 	&	$	+0.05 	$	&	$	+26 	$	&	424 	&	317 	&	$	<	555 	$	&	184 	/	164	\\
	&	7	&	5.56	&	5.43 	&	0.35 	&	0.11 	&	$	-0.36 	$	&	$	-397 	$	&	1056 	&	317 	&	$	<	1500 	$	&	44 	/	36	\\
	&	8	&	5.56	&	4.84 	&	0.62 	&	0.11 	&	$	-1.15 	$	&	$	-2169 	$	&	1862 	&	317 	&	$	<	4057 	$	&	43 	/	24	\\
	&	9	&	5.56	&	5.60 	&	0.24 	&	0.11 	&	$	+0.14 	$	&	$	+111 	$	&	714 	&	317 	&	$	<	892 	$	&	50 	/	67	\\
	&	10	&	5.56	&	5.28 	&	0.29 	&	0.11 	&	$	-0.90 	$	&	$	-826 	$	&	863 	&	317 	&	$	<	1745 	$	&	101 	/	105	\\
	&	11	&	5.56	&	5.54 	&	0.33 	&	0.11 	&	$	-0.07 	$	&	$	-67 	$	&	987 	&	317 	&	$	<	1104 	$	&	46 	/	41	\\
	&	13	&	5.56	&	6.50 	&	0.81 	&	0.11 	&	$	+1.15 	$	&	$	+2828 	$	&	2444 	&	317 	&	$	<	5292 	$	&	30 	/	40	\\
	&	14	&	5.56	&	6.03 	&	0.42 	&	0.11 	&	$	+1.09 	$	&	$	+1410 	$	&	1259 	&	317 	&	$	<	2708 	$	&	45 	/	40	\\ \hline
A2029	&	1	&	7.73	&	7.25 	&	0.21 	&	0.11 	&	$	-2.05 	$	&	$	-1443 	$	&	623 	&	323 	&	$	<	2145 	$	&	103 	/	117	\\
	&	2	&	7.73	&	7.62 	&	0.11 	&	0.11 	&	$	-0.73 	$	&	$	-338 	$	&	328 	&	323 	&	$	<	799 	$	&	209 	/	182	\\
	&	3	&	7.73	&	7.76 	&	0.32 	&	0.11 	&	$	+0.08 	$	&	$	+76 	$	&	949 	&	323 	&	$	<	1078 	$	&	106 	/	83	\\
	&	4	&	7.73	&	7.49 	&	0.15 	&	0.11 	&	$	-1.28 	$	&	$	-711 	$	&	450 	&	323 	&	$	<	1265 	$	&	134 	/	119	\\ \hline
A2255	&	1	&	8.06	&	8.14 	&	0.21 	&	0.11 	&	$	+0.34 	$	&	$	+242 	$	&	625 	&	324 	&	$	<	946 	$	&	199 	/	247	\\
	&	2	&	8.06	&	8.28 	&	0.17 	&	0.11 	&	$	+1.12 	$	&	$	+673 	$	&	506 	&	324 	&	$	<	1275 	$	&	259 	/	232	\\
	&	3	&	8.06	&	8.28 	&	0.41 	&	0.11 	&	$	+0.52 	$	&	$	+667 	$	&	1231 	&	324 	&	$	<	1940 	$	&	110 	/	112	\\
       &	4	&	8.06	&	7.35 	&	0.26 	&	0.11 	&	$	-2.56 	$	&	$	-2141 	$	&	772 	&	324 	&	$	<	2978 	$	&	128 	/	136	\\ \hline
A2142	&	1	&	9.09	&	8.94 	&	0.13 	&	0.11 	&	$	-0.88 	$	&	$	-444 	$	&	386 	&	327 	&	$	<	951 	$	&	304 	/	292	\\
	&	2	&	9.09	&	8.99 	&	0.10 	&	0.11 	&	$	-0.69 	$	&	$	-310 	$	&	304 	&	327 	&	$	<	757 	$	&	232 	/	293	\\
	&	4	&	9.09	&	8.81 	&	0.30 	&	0.11 	&	$	-0.87 	$	&	$	-843 	$	&	912 	&	327 	&	$	<	1812 	$	&	89 	/	90	\\
	&	5	&	9.09	&	9.14 	&	0.06 	&	0.11 	&	$	+0.36 	$	&	$	+137 	$	&	192 	&	327 	&	$	<	516 	$	&	451 	/	485	\\
	&	6	&	9.09	&	9.21 	&	0.09 	&	0.11 	&	$	+0.87 	$	&	$	+370 	$	&	269 	&	327 	&	$	<	794 	$	&	504 	/	557	\\
	&	7	&	9.09	&	8.60 	&	0.53 	&	0.11 	&	$	-0.91 	$	&	$	-1482 	$	&	1595 	&	327 	&	$	<	3110 	$	&	76 	/	72	\\
	&	8	&	9.09	&	9.27 	&	0.61 	&	0.11 	&	$	+0.29 	$	&	$	+544 	$	&	1835 	&	327 	&	$	<	2408 	$	&	40 	/	44	\\
	&	9	&	9.09	&	9.13 	&	0.30 	&	0.11 	&	$	+0.13 	$	&	$	+126 	$	&	887 	&	327 	&	$	<	1071 	$	&	118 	/	101	\\
	&	10	&	9.09	&	8.88 	&	0.30 	&	0.11 	&	$	-0.66 	$	&	$	-634 	$	&	908 	&	327 	&	$	<	1599 	$	&	125 	/	134	\\
	&	11	&	9.09	&	8.82 	&	0.39 	&	0.11 	&	$	-0.67 	$	&	$	-806 	$	&	1158 	&	327 	&	$	<	2010 	$	&	45 	/	47	\\
	&	13	&	9.09	&	8.70 	&	0.49 	&	0.11 	&	$	-0.79 	$	&	$	-1175 	$	&	1460 	&	327 	&	$	<	2670 	$	&	38 	/	44	\\
	&	14	&	9.09	&	9.69 	&	0.89 	&	0.11 	&	$	+0.67 	$	&	$	+1805 	$	&	2677 	&	327 	&	$	<	4502 	$	&	41 	/	46	\\ \hline
\end{longtable}

\begin{figure}
\begin{center}
\rotatebox{0}{\scalebox{0.3}{\includegraphics{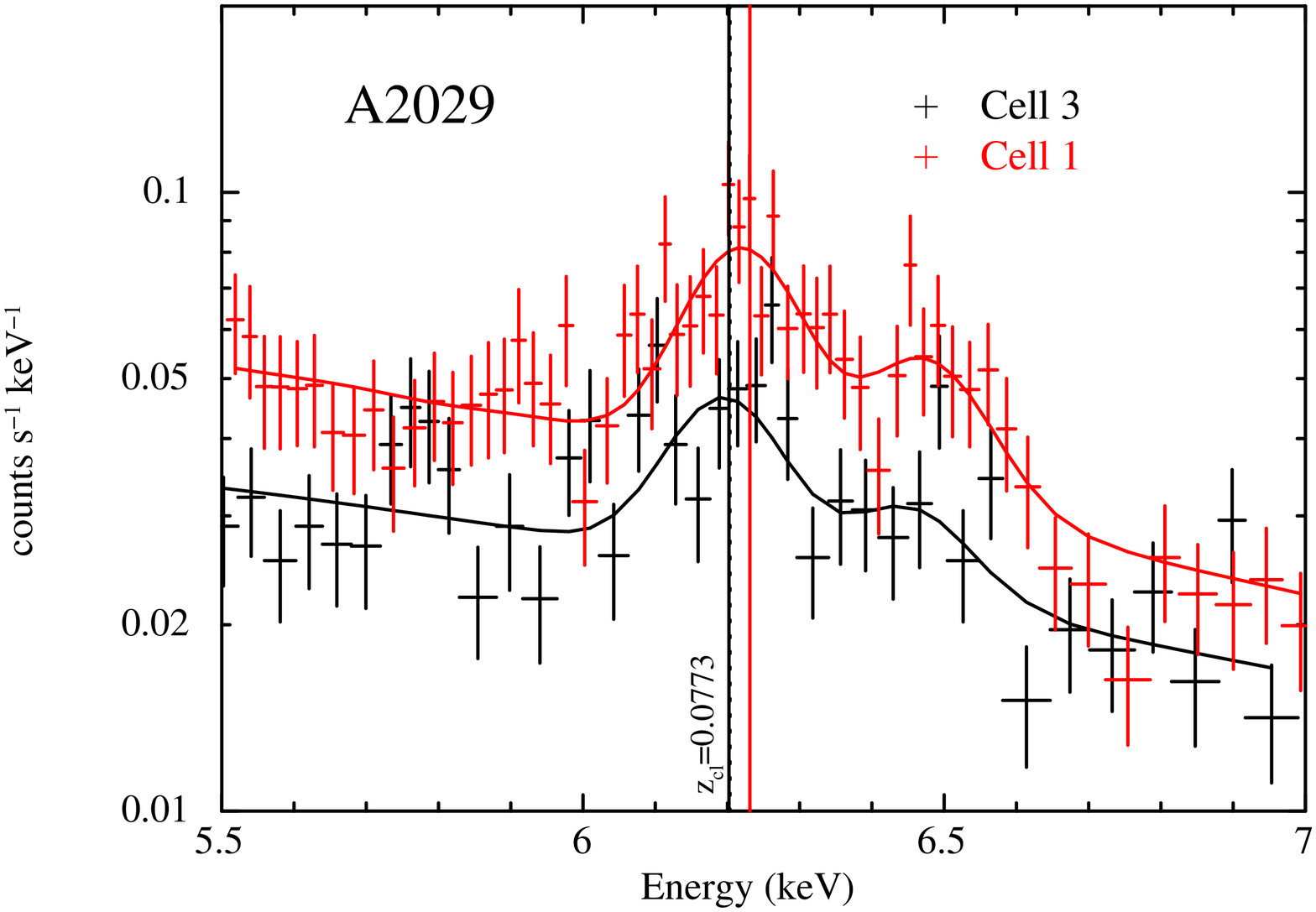}}}

\rotatebox{0}{\scalebox{0.3}{\includegraphics{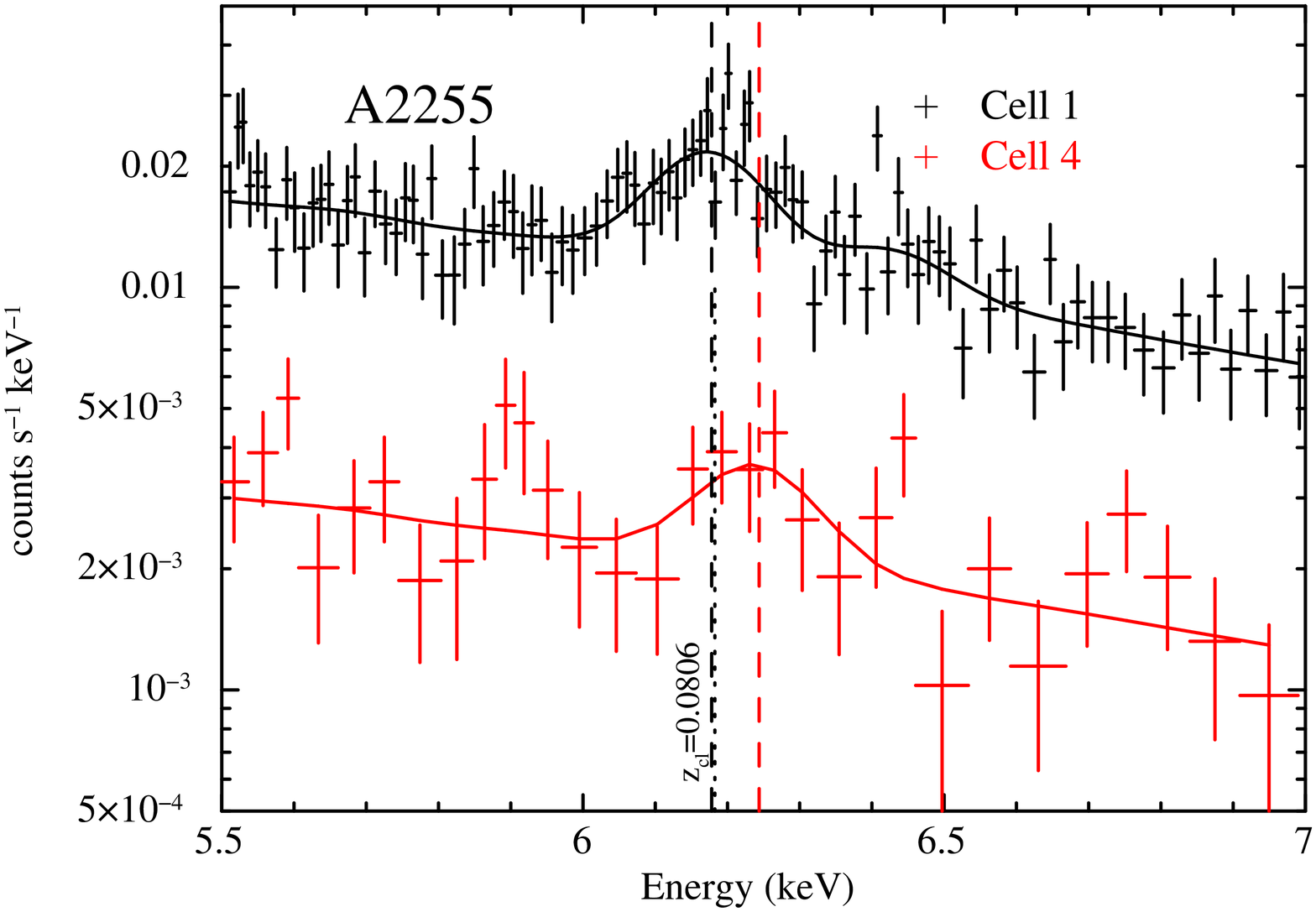}}}

\rotatebox{0}{\scalebox{0.3}{\includegraphics{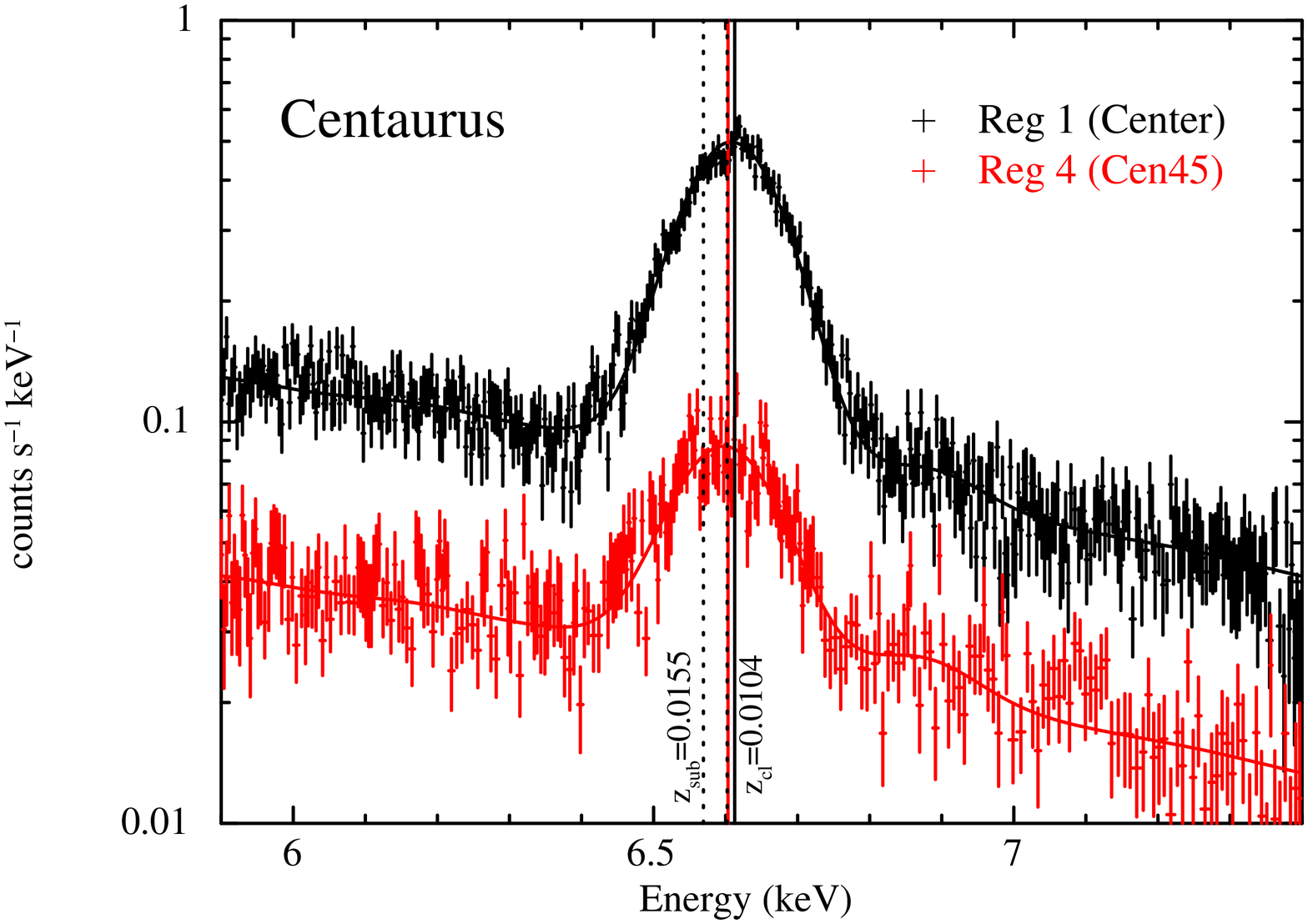}}}
\end{center}
\caption{XIS spectra of A2029 (top), and
  A2255 (middle), and Centaurus (bottom). In each panel, the spectra of two cells
  inside each cluster (crosses) and the best-fit Gaussian models (the
  solid curves) are shown. The best-fit centroid energies of the
  He-like Fe-K line complex are indicated by the vertical dashed 
  lines. The line centroid energies expected from the cluster redshifts and that from the subcluster in Centaurus are shown by the dotted lines. }\label{fig5}
\end{figure}

Table~\ref{tab5} gives the summary of redshift and radial velocity of
gas bulk motion for cells with $\gtrsim 2\sigma$ significance: A2029 cell
No.~2 and A2255 cell No.~4.  For reference, the values for cells
without any significant motion, A2029 cell No.~3 and A2255 cell No.~1,
are also listed.  Figure~\ref{fig5} shows the XIS spectra of the Fe-K
line emission from these objects.

As shown in Table~\ref{tab5} and Figure~\ref{fig5}, we found that the
ICM in the Cen45 subcluster has a redshift consistent with that of the
main cluster $z_{\rm cl}=0.0104$ but is different from that of the
subcluster in the optical band $z_{\rm sub}=0.0155$ at the $3.9\sigma$
significance. The origin of the redshift difference is discussed in \S\ref{subsubsec:cen}.

\begin{table*}
\tbl{ICM redshift and the radial velocity of gas bulk motion}{
\begin{tabular}{lllllll} \hline 
Cluster & Cell & $z_{\rm cl}~(10^{-2})$ & $z~(10^{-2})$ & $|\Delta z/\sigma_{\rm z,tot}|$$^{\mathrm{a}}$ & $\Delta v$ & $s$$^{\mathrm{b}}$ \\ 
 &  &  &  & ${\rm (km\,s^{-1})}$ & ${\rm (km\,s^{-1})}$\\ \hline
A2029 & 1 & 7.73 & $7.25\pm0.23$ & 2.1 & $-1443\pm702$ & 1403  \\
            & 3$^{\mathrm{c}}$  & 7.73 & $7.76\pm0.33$  & 0.1 & $76\pm1002$ & 1403\\ \hline 
A2255 & 4 & 8.06 & $7.35\pm0.28$ & 2.6 & $-2141\pm837$ &1248 \\ 
           & 1$^{\mathrm{c}}$ & 8.06 & $8.14\pm0.23$  & 0.3 & $242\pm704$ & 1248 \\\hline  
Cen45  & 4 & 1.04 & $1.12\pm0.11$ & 0.7 &  $+227\pm332$ & 880 \\ 
            & 4 & 1.55$^{\mathrm{d}}$ & $1.12\pm0.11$ & 3.9 &  $-1303\pm332$ & 880 \\ \hline
\end{tabular}}\label{tab5}
\begin{tabnote}
The $1\sigma$ errors including the statistical and systematic uncertainties are quoted.
$^{\mathrm{a}}$ Significance of the redshift difference, $\Delta z$, from the XIS spectral fit divided by the standard deviation including the statistical and systematic uncertainties.
$^{\mathrm{b}}$ Sound velocity of the ICM (see \S\ref{subsec:mass_estimate}).
$^{\mathrm{c}}$ Reference cell.
$^{\mathrm{d}}$ The optical redshift of the Cen45 subcluster.
\end{tabnote}
\end{table*}

\section{Discussion}
On the basis of the {\it Suzaku}/XIS spectral analysis and careful
assessment of the systematic error of the energy scale, we measured
the Doppler shift of He-like Fe-K line emission to search for
the ICM bulk motion in the central regions of eight nearby clusters. A
signature of gas bulk motion on the order of $1000-2000~{\rm
  km\,s^{-1}}$ was discovered in A2029 and A2255.  On the contrary, no
significant bulk motion was detected in five clusters. In
\S\ref{subsec:velocity_structure}, we compare the velocity
distributions of ICM measured in \S\ref{sec:analysis} with those of
member galaxies for individual clusters to infer the dynamical
evolution of clusters. In \S\ref{subsec:mass_estimate}, we discuss the
impact of ICM bulk motion on the estimation of cluster gravitational
mass under the assumption of hydrostatic equilibrium.

\subsection{Velocity structures of gas and galaxies in individual clusters}\label{subsec:velocity_structure}
\subsubsection{The Centaurus cluster}\label{subsubsec:cen}
Optical observations of the Centaurus cluster show that this cluster
has a bimodal velocity distribution of galaxies: one is associated
with the main cluster centered on NGC 4696, and the other with the
Cen45 subcluster centered on NGC~4709 \citep{1986MNRAS.222..427L}.
The Cen45 subcluster has a large velocity difference of $+1500~{\rm
  km\,s^{-1}}$ relative to the main cluster.  Previous {\it ASCA}
observations show a temperature excess in the subcluster region
\citep{2001ApJ...561L.165F}.

The present analysis using the latest XIS energy response yielded the
redshift values for the three pointings centered on the main cluster
(i.e., Centaurus cluster, Offset1, and Offset 2), which is consistent
with earlier results by \citet{2007PASJ...59S.351O}. This result
confirms that the bulk velocity does not largely exceed the sound
velocity in the cluster core.

As shown in Table~\ref{tab3}, the fitted redshift of Cen45
($0.0116\pm0.0005$) is higher than that of the center
($0.00975\pm0.00007$); however, the difference is within the range of
their systematic errors.  Recently, \citet{2013MNRAS.435.3221W}
confirmed the temperature excess near Cen45 on the basis of the XMM
data. They modeled {\it XMM} spectra of Cen45 by superposition of two
emission components from the main cluster and the Cen45 subcluster. We
thus fitted the XIS spectra from a $r=7\arcmin$ circle centered on
NGC~4709 by including their best-fit model for the $z=0.0104$ main
cluster component in our background model to obtain the Cen45 redshift
as $0.0105\pm0.0011$. Therefore the redshift difference relative to
the optical subcluster is significant at the $4\sigma$ level.

The 90\% upper limit on the radial velocity of the ICM in the Cen45
subcluster was estimated as $|\Delta v| < 760 ~{\rm km\,s^{-1}}$. This
value is twice smaller than that of the galaxies, at $+1500~{\rm
  km\,s^{-1}}$, suggesting the presence of an offset between the mass
centroids of the ICM and galaxy distributions along the sightline.  A
spatial segregation between a dissipationless stellar component and
fluid-like X-ray emitting plasma has been discovered in on-going
mergers such as the Bullet cluster \citep{2006ApJ...648L.109C} and
A2744 \citep{2011MNRAS.417..333M}.  Thus the above result supports
that the Centaurus cluster has experienced a subcluster merger along
the line of sight and gas near Cen45 has been heated by the strong
shock \citep{2013MNRAS.435.3221W,1999ApJ...520..105C}.

\subsubsection{The Perseus cluster}
The Perseus cluster is a nearby relaxed cluster with a cool core at
the center. This cluster is one of the brightest in the sky and has an
intense Fe line emission. The velocity dispersion of member galaxies
is approximately $1000~{\rm km\,s^{-1}}$
\citep{1983AJ.....88..697K}. The XIS spectra analysis of $4\times4$
cells in the center showed that the bulk velocity does not
significantly exceed the sound velocity of the ICM. The 90\% limit on
the radial velocity is $|\Delta v|< 830~{\rm km\,s^{-1}}$, which is
comparable to that obtained for the Centaurus cluster and also agrees
with an in-depth analysis of {\it Suzaku's} multiple pointing data
\citep{tamura14}.
 
\subsubsection{A2199}
A2199 is a nearby relaxed cluster with smooth and symmetric
gas morphology. The cD galaxy hosts a radio jet and its interaction
with the surrounding gas was shown by the {\it ROSAT} observations
\citep{1998ApJ...493...73O}. \citet{2006MNRAS.371L..65S} found a 
density jump in the cluster center, implying a Mach number $\sim 1.5$
shock. In addition, deep {\it Chandra} observations showed the
presence of gas sloshing due to a minor merger
\citep{2013ApJ...775..117N}.

From the XIS analysis of the central region, we found a sign of large
bulk velocity $\Delta v \sim +700-+2000~{\rm km\,s^{-1}}$ in cell
Nos.~2, 7, 9, and 14, suggesting that the gas motion is associated
with the infalling subcluster as indicated by {\it Chandra}
data. Considering the calibration error, the significance of the
detection of ICM motion is $1.6-1.8\sigma$, thus needs to be improved
by follow-up observations.

\subsubsection{A3667}
A3667 is a merging cluster with an elongated X-ray morphology. Cold
fronts were detected by {\it Chandra} observations
\citep{2001ApJ...549L..47V}. The spatial and redshift distributions of
member galaxies are bimodal \citep{2009ApJ...693..901O}. Their mean
radial velocities are separated by $500~{\rm km\,s^{-1}}$ and the
maximum velocity gradient across the cluster is $2800~{\rm
  km\,s^{-1}}$. This cluster exhibits two radio relics and a
temperature discontinuity, suggesting that A3667 has recently
experienced a strong merger shock
\citep{1999ApJ...518..603R,2012PASJ...64...49A}.

The redshifts of Fe-K lines measured for A3667 Center and NW
(Fig.~\ref{fig1}, Table~\ref{tab3}) are consistent with the optical
values of $z=0.0555$ and $0.0567$ for the two regions, respectively.
Given the measurement uncertainties, however, the velocity difference
between the two regions is not significant, resulting in 90\% limit on
the radial velocity $< 2800~{\rm km\,s^{-1}}$ from the large-scale
measurements.  Although a large bulk velocity of $+1510\pm960~{\rm
  km\,s^{-1}}$ is marginally detected in cell No.~4, we did not find
any significant bulk motion in the cluster center (Fig.~\ref{fig2},
Table~\ref{tab4}), giving the 90\% limit $|\Delta v|< 3500~{\rm
  km\,s^{-1}}$ from the small-scale measurements.

\subsubsection{A133}
A133 is an X-ray luminous cluster hosting a cD cluster and a radio
relic. {\it Chandra} observations showed that the central cool core is
irregular and extends to the northwest of the cD galaxy
\citep{2002ApJ...575..764F}. From {\it XMM} observations, a cold front
was detected to the southeast of the cluster core and an upper limit
to the velocity of the core relative to the rest of the cluster was
estimated as $<230~{\rm km\,s^{-1}}$ in the context of the
Kelvin-Helmholtz instability around the core
\citep{2004ApJ...616..157F}. Using a wavelet analysis of galaxy
positions, \citet{2010ApJ...722..825R} found three distinct
substructures in the cluster. They also identified a concentration of
galaxies to the southwest, whose radial velocity histogram is peaked
at approximately $-1000~{\rm km\,s^{-1}}$.

We did not find any significant variation in ICM velocity among the
four pointing regions and placed the upper limit on the ICM radial
velocity $|\Delta v| < 3300~{\rm km\,s^{-1}}$ at the 90\% confidence
level.

\subsubsection{A2029}
A2029 is an X-ray bright, relaxed cluster, hosting a cool core at the
center. A high gas temperature of 7~keV and velocity dispersion of
$1430~{\rm km\,s^{-1}}$ \citep{1979ApJ...231..659D} indicate a large
gravitational mass.  On the basis of deep {\it Chandra} observations,
\citet{2013ApJ...773..114P} reported that there is a spiral structure
in the core region produced by sloshing motion of the ICM.

In the southeast region (cell No.~1), we detected a line-of-sight gas
motion in excess of the systematic error. The analysis of {\it Suzaku}
data yielded the velocity difference $\Delta v = -1440 \pm 700~{\rm
  km\,s^{-1}}$. Thus, the detection is significant at the $2.1\sigma$
level.  The measured gas velocity is likely to be larger than that
expected for the sloshing motion in the cluster, at $\sim 300~{\rm
  km\,s^{-1}}$ \citep{2013ApJ...773..114P}.  The gas bulk motion will
introduce a systematic uncertainty in the cluster mass, which will be
discussed in the next subsection.

\subsubsection{A2255}
A2255 is a nearby rich cluster with an elongated X-ray morphology
\citep{1995ApJ...446..583B}. The asymmetric temperature distribution
\citep{2006MNRAS.367.1409S} and the presence of radio halo and relics
\citep{2009A&A...507..639P} indicate that the cluster has undergone a
recent merger. In the optical, A2255 has a large velocity dispersion
of $1200~{\rm km\,s^{-1}}$ and several substructures approximately
$10\arcmin-15\arcmin$ from the center. In particular, the NW and SE
subclusters have a radial velocity of $-920$ and $+870~{\rm
  km\,s^{-1}}$, respectively \citep{2003ApJS..149...53Y}, which is
also indicative of mergers.

From the Gaussian fitting to the Fe-line spectra in $2\times2$ cells,
the redshift of the NW region (cell No.~4) was obtained as
$z=0.073\pm0.003$. This value is lower than that of optical redshift
of the main cluster $z_{\rm cl}=0.0806$ and is rather close to that of
the NW subcluster $0.077$. From the XIS analysis, the velocity
difference between the main cluster and NW subcluster was determined
as $|\Delta v|\sim 2100~{\rm km\,s^{-1}}$, which is comparable to that
of A2256 reported by \citet{2011PASJ...63S1009T}.  The present results
support the idea such that the gas and galaxies are moving together as
a single substructure within the cluster potential.

Considering the the accuracy of the XIS energy-scale, the detection of
gas motion in the NW region is at the $2.6\sigma$ significance.  A
further examination, however, must await follow-up X-ray observations
of subcluster regions.

\subsubsection{A2142}
A2142 is a nearby merging cluster and is the first object in which
cold fronts were identified by {\it Chandra}
\citep{2000ApJ...541..542M}. This cluster is hot and X-ray luminous
and has two cD galaxies near the center. The line-of-sight velocities
of these galaxies differ by $1600~{\rm km\,s^{-1}}$
\citep{1995AJ....110...32O}, suggesting that the cluster is not
dynamically relaxed. \citet{2011ApJ...741..122O} utilized the spatial
and spectral data of member galaxies to identify many group-scale
substructures in A2142. Two of the subgroups denoted as S1 and S2 are
located at $1\arcmin.8$ and $10\arcmin$ off of the cluster center. S1
is centered on the secondary cD galaxy, whereas S2 has a mean velocity
of approximately $1700~{\rm km\,s^{-1}}$ relative to the main cluster
mass peak.

The redshifts derived from the XIS spectral analysis are consistent
with the optical values within the current XIS calibration
uncertainty.  The S1 and S2 substructures as well as the main cluster
peak are inside the XIS field of view. No significant bulk motion
associated with the optical substructures was detected by {\it
  Suzaku}, giving the 90\% upper limit $|\Delta v|< 4200~{\rm
  km\,s^{-1}}$.

\subsection{Impact of Gas Bulk Motions on Cluster Mass Estimation}\label{subsec:mass_estimate}
If significant bulk motions exist, a contribution of nonthermal
pressure must be considered in estimating the cluster mass.
Table~\ref{tab5} shows a comparison of the ICM bulk velocity with the
sound velocity $s=(\gamma kT/\mu m_p)^{1/2}$ in the cluster center in
which the sign of bulk motions was observed. Here, $\gamma$ is the
adiabatic index and $m_p$ is the proton mass. Their bulk velocity is
marginally larger than the sound velocity.

On the basis of the {\it Suzaku} observations of the Centaurus
cluster, \citet{2007PASJ...59S.351O} investigated the possible impact
of nonthermal pressure on the cluster mass estimate assuming simply
that the gas is rotating at a circular velocity $\sigma_r$.  The
balance against the gravitational pull at a radius $r$ is given by
\begin{eqnarray}
-\frac{GM(r)\rho_{\rm gas}}{r^2} &=& \frac{\partial P_{\rm gas}}{\partial r} - \frac{f\rho_{\rm gas}\sigma_r^2}{r} 
\sim \frac{\partial }{\partial r}P_{\rm gas}(1+f\beta_r), \label{eq:HS} \\ 
\beta_r &=& \frac{\mu m_p \sigma_r^2}{kT}\sim 1.07\left(\frac{\mu}{0.63}\right)\left(\frac{\sigma_r}{700{\rm km\,s^{-1}}}\right)^2\left(\frac{kT}{3{\rm keV}}\right)^{-1}, \label{eq:beta}
\end{eqnarray}
where $M(r)$ is the cluster mass interior to $r$, $\rho_{\rm gas}$ and
$P_{\rm gas}$ are the gas density and thermal pressure, and $f$ is a
fraction of the ICM that is rotating. Thus the total cluster mass
should be higher than the hydrostatic mass by a factor of $(1 +
f\beta_r)$ because of the presence of additional pressure support.

We calculate the mass correction factor $(1+ f\beta_r)$ assuming
$\sigma_r$ to be given by the observed $\Delta v$
(Table~\ref{fig5}). Following the same manner as in \citet{tamura14},
$f$ is approximated by the emission fraction from an interesting cell.
As a result, the mass correction factor is estimated to be 1.9, 2.4
for A2029 cell No.~1, A2255 cell No.~4, respectively.  Therefore, the
results indicate that the nonthermal pressure support cannot be
neglected in some regions of both mergers and relaxed clusters as
suggested by the numerical simulations. These are, however, crude
estimations and more accurate information on the spatial velocity
structure is required to improve the precision of the mass
determination.

The SXS microcalorimeter on board the {\it ASTRO-H} satellite is a
non-dispersive spectrometer and enables high-resolution (5~eV)
observations in the X-ray regime \citep{2014SPIE.9144E..2AM}. SXS can
measure the Doppler shift and broadening of iron lines caused by
kinetic gas motions to an accuracy of $100~{\rm km\,s^{-1}}$, which
plays an critical role in revealing the dynamics of clusters (e.g.,
\cite{2013ApJ...777..137N,2012RAA....12..973O,2014arXiv1412.1176K}).

\section{Summary}
We used the {\it Suzaku} satellite to search for gas bulk motion in
eight nearby clusters with a variety of X-ray morphologies. From the
model fitting to the Fe-K line spectra, the Doppler shift was measured
to high accuracy. The velocity structure of gas was compared with that
of member galaxies in the optical band to study the dynamical states
of clusters. This study is also important as a pilot survey for the
future high-resolution spectroscopy with the X-ray microcalorimeter
onboard {\it ASTRO-H}.

In the cores of Centaurus and Perseus, we confirmed that the bulk
velocity does not largely exceed the sound velocity. For the Cen45
subcluster, we found that the upper limit on the ICM radial velocity
($|\Delta v|<760~{\rm km\,s^{-1}}$) is significantly smaller than that
of galaxies, suggesting that there is an offset between the gas and
galaxy distributions along the line of sight due to the subcluster
merger.
 
In the cool-core cluster A2199, we found a sign of large bulk
velocity, however, the significance is not sufficiently high to claim
the detection.  For two merging clusters with cold fronts, A2142 and
A3667, and a cluster with an irregular cool core, A133, no significant
gas motion was detected, yielding the 90\% upper limit of
$3000-4000~{\rm km\,s^{-1}}$.

A sign of large bulk velocity in excess of the XIS calibration
uncertainty was found in the cool-core cluster A2029, and in the A2255
subcluster. We examined the impact of nonthermal pressure support in
the cluster mass estimation to find it is not negligible in some
regions of both relaxed and merging clusters as predicted by numerical
simulations. To improve the significance of the detection, however, a
further examination by follow-up observations is required.

We expect that upcoming high-resolution observations with {\it
  ASTRO-H} will determine the ICM velocity structure more accurately
and will clarify the contribution of nonthermal pressure support in
the cluster mass estimation.

\begin{ack}
  We are grateful to the Suzaku team members for satellite operation
  and instrumental calibration. We also thank the anonymous referee for useful comments. This was supported in part by JSPS KAKENHI grant 25400231, 25247028 (NO).
\end{ack}

\end{document}